\documentstyle[twoside,epsfig]{article}

\catcode`\@=11
\long\def\@makefntext#1{
\protect\noindent \hbox to 3.2pt {\hskip-.9pt
$^{{\eightrm\@thefnmark}}$\hfil}#1\hfill}               

\def\@makefnmark{\hbox to 0pt{$^{\@thefnmark}$\hss}}    

\def\ps@myheadings{\let\@mkboth\@gobbletwo
\def\@oddhead{\hbox{}
\rightmark\hfil\eightrm\thepage}
\def\@oddfoot{}\def\@evenhead{\eightrm\thepage\hfil
\leftmark\hbox{}}\def\@evenfoot{}
\def\sectionmark##1{}\def\subsectionmark##1{}}

\oddsidemargin=\evensidemargin
\newcounter{sectionc}\newcounter{subsectionc}\newcounter{subsubsectionc}
\renewcommand{\section}[1] {\vspace{12pt}\addtocounter{sectionc}{1}
\setcounter{subsectionc}{0}\setcounter{subsubsectionc}{0}\noindent
       {\tenbf\thesectionc. #1}\par\vspace{5pt}}
\renewcommand{\subsection}[1] {\vspace{12pt}\addtocounter{subsectionc}{1}
       \setcounter{subsubsectionc}{0}\noindent
       {\bf\thesectionc.\thesubsectionc. {\kern1pt \bfit #1}}\par\vspace{5pt}}
\renewcommand{\subsubsection}[1] {\vspace{12pt}\addtocounter{subsubsectionc}{1}
       \noindent{\tenrm\thesectionc.\thesubsectionc.\thesubsubsectionc.
       {\kern1pt \tenit #1}}\par\vspace{5pt}}
\newcommand{\nonumsection}[1] {\vspace{12pt}\noindent{\tenbf #1}
       \par\vspace{5pt}}

\newcounter{appendixc}
\newcounter{subappendixc}[appendixc]
\newcounter{subsubappendixc}[subappendixc]
\renewcommand{\thesubappendixc}{\Alph{appendixc}.\arabic{subappendixc}}
\renewcommand{\thesubsubappendixc}
       {\Alph{appendixc}.\arabic{subappendixc}.\arabic{subsubappendixc}}

\renewcommand{\appendix}[1] {\vspace{12pt}
       \refstepcounter{appendixc}
       \setcounter{figure}{0}
       \setcounter{table}{0}
       \setcounter{lemma}{0}
       \setcounter{theorem}{0}
       \setcounter{corollary}{0}
       \setcounter{definition}{0}
       \setcounter{equation}{0}
       \renewcommand{\thefigure}{\Alph{appendixc}.\arabic{figure}}
       \renewcommand{\thetable}{\Alph{appendixc}.\arabic{table}}
       \renewcommand{\theappendixc}{\Alph{appendixc}}
       \renewcommand{\thelemma}{\Alph{appendixc}.\arabic{lemma}}
       \renewcommand{\thetheorem}{\Alph{appendixc}.\arabic{theorem}}
       \renewcommand{\thedefinition}{\Alph{appendixc}.\arabic{definition}}
       \renewcommand{\thecorollary}{\Alph{appendixc}.\arabic{corollary}}
       \noindent{\tenbf Appendix \theappendixc #1}\par\vspace{5pt}}
\newcommand{\subappendix}[1] {\vspace{12pt}
       \refstepcounter{subappendixc}
       \noindent{\bf Appendix \thesubappendixc. {\kern1pt \bfit #1}}
       \par\vspace{5pt}}
\newcommand{\subsubappendix}[1] {\vspace{12pt}
       \refstepcounter{subsubappendixc}
       \noindent{\rm Appendix \thesubsubappendixc. {\kern1pt \tenit #1}}
       \par\vspace{5pt}}

\newcommand{\textlineskip}{\baselineskip=13pt}
\newcommand{\smalllineskip}{\baselineskip=10pt}

\def\eightcirc{
\begin{picture}(0,0)
\put(4.4,1.8){\circle{6.5}}
\end{picture}}
\def\eightcopyright{\eightcirc\kern2.7pt\hbox{\eightrm c}}

\newcommand{\copyrightheading}[1]
       {\vspace*{-2.5cm}\smalllineskip{\flushleft
       {\footnotesize International Journal of Modern Physics C #1}\\
       {\footnotesize $\eightcopyright$\, World Scientific Publishing
        Company}\\
        }}

\def\abstracts#1#2#3{{
       \centering{\begin{minipage}{4.5in}\footnotesize\baselineskip=10pt
       \parindent=0pt #1\par
       \parindent=15pt #2\par
       \parindent=15pt #3
       \end{minipage}}\par}}

\def\keywords#1{{
       \centering{\begin{minipage}{4.5in}\footnotesize\baselineskip=10pt
       {\footnotesize\it Keywords}\/: #1
       \end{minipage}}\par}}

\newcommand{\bibit}{\nineit}
\newcommand{\bibbf}{\ninebf}
\renewenvironment{thebibliography}[1]
       {\frenchspacing
        \ninerm\baselineskip=11pt
        \begin{list}{\arabic{enumi}.}
       {\usecounter{enumi}\setlength{\parsep}{0pt}
        \setlength{\leftmargin 12.7pt}{\rightmargin 0pt} 
        \setlength{\itemsep}{0pt} \settowidth
       {\labelwidth}{#1.}\sloppy}}{\end{list}}

\newcounter{itemlistc}
\newcounter{romanlistc}
\newcounter{alphlistc}
\newcounter{arabiclistc}

\newcommand{\fcaption}[1]{
       \refstepcounter{figure}
       \setbox\@tempboxa = \hbox{\footnotesize Fig.~\thefigure. #1}
       \ifdim \wd\@tempboxa > 5in
          {\begin{center}
       \parbox{5in}{\footnotesize\smalllineskip Fig.~\thefigure. #1}
           \end{center}}
       \else
            {\begin{center}
            {\footnotesize Fig.~\thefigure. #1}
             \end{center}}
       \fi}

\newcommand{\tcaption}[1]{
       \refstepcounter{table}
       \setbox\@tempboxa = \hbox{\footnotesize Table~\thetable. #1}
       \ifdim \wd\@tempboxa > 5in
          {\begin{center}
       \parbox{5in}{\footnotesize\smalllineskip Table~\thetable. #1}
           \end{center}}
       \else
            {\begin{center}
            {\footnotesize Table~\thetable. #1}
             \end{center}}
       \fi}

\def\@citex[#1]#2{\if@filesw\immediate\write\@auxout
       {\string\citation{#2}}\fi
\def\@citea{}\@cite{\@for\@citeb:=#2\do
       {\@citea\def\@citea{,}\@ifundefined
       {b@\@citeb}{{\bf ?}\@warning
       {Citation `\@citeb' on page \thepage \space undefined}}
       {\csname b@\@citeb\endcsname}}}{#1}}

\newif\if@cghi
\def\cite{\@cghitrue\@ifnextchar [{\@tempswatrue
       \@citex}{\@tempswafalse\@citex[]}}
\def\citelow{\@cghifalse\@ifnextchar [{\@tempswatrue
       \@citex}{\@tempswafalse\@citex[]}}
\def\@cite#1#2{{$\null^{#1}$\if@tempswa\typeout
       {IJCGA warning: optional citation argument
       ignored: `#2'} \fi}}

\def\pmb#1{\setbox0=\hbox{#1}
       \kern-.025em\copy0\kern-\wd0
       \kern.05em\copy0\kern-\wd0
       \kern-.025em\raise.0433em\box0}

\def\fnt#1#2{\footnotetext{\kern-.3em
       {$^{\mbox{\scriptsize #1}}$}{#2}}}

\def\ps@myheadings{%
   \let\@oddfoot\@empty\let\@evenfoot\@empty
   \def\@evenhead{\slshape\leftmark\hfil}
   \def\@oddhead{\hfil{\slshape\rightmark}}
   \let\@mkboth\@gobbletwo
   \let\sectionmark\@gobble
   \let\subsectionmark\@gobble
   }
\font\tenrm=cmr10
\font\tenit=cmti10
\font\tenbf=cmbx10
\font\bfit=cmbxti10 at 10pt
\font\ninerm=cmr9
\font\nineit=cmti9
\font\ninebf=cmbx9
\font\eightrm=cmr8

\def\qed{\hbox{${\vcenter{\vbox{                    
  \hrule height 0.4pt\hbox{\vrule width 0.4pt height 6pt
  \kern5pt\vrule width 0.4pt}\hrule height 0.4pt}}}$}}

\def\bsc{{\sc a\kern-6.4pt\sc a\kern-6.4pt\sc a}}       
\def\bflatex{\bf L\kern-.30em\raise.3ex\hbox{\bsc}\kern-.14em
T\kern-.1667em\lower.7ex\hbox{E}\kern-.125em X}

\begin{document}
\setlength{\textheight}{7.7truein}  

\thispagestyle{empty}


\normalsize\textlineskip

\setcounter{page}{1}

\copyrightheading{}                     

\vspace*{0.88truein}
\centerline{\bf VECTOR OPINION DYNAMICS}
\vspace*{0.035truein}
\centerline{\bf IN A BOUNDED CONFIDENCE CONSENSUS MODEL}
\vspace*{0.37truein}
\centerline{\footnotesize SANTO FORTUNATO}
\baselineskip=12pt
\centerline{\footnotesize\it Fakult\"at f\"ur Physik, Universit\"at Bielefeld}
\baselineskip=10pt
\centerline{\footnotesize\it D-33501 Bielefeld, Germany $\&$}
\baselineskip=10pt
\centerline{\footnotesize\it School of Informatics, Indiana University}
\baselineskip=10pt
\centerline{\footnotesize\it Bloomington, IN 47408, USA $\&$}
\baselineskip=10pt
\centerline{\footnotesize\it Dipartimento di Fisica e Astronomia and INFN sezione di Catania}
\baselineskip=10pt
\centerline{\footnotesize\it Universita' di Catania, Catania I-95123, Italy}
\centerline{\footnotesize\it E-mail: santo@indiana.edu}

\vspace*{15pt}          
\centerline{\footnotesize VITO LATORA}
\baselineskip=12pt
\centerline{\footnotesize\it Dipartimento di Fisica e Astronomia and INFN sezione di Catania}
\baselineskip=10pt
\centerline{\footnotesize\it Universita' di Catania, Catania I-95123, Italy}
\centerline{\footnotesize\it E-mail: Vito.Latora@ct.infn.it}
\vspace*{15pt}          
\centerline{\footnotesize ALESSANDRO PLUCHINO}
\baselineskip=12pt
\centerline{\footnotesize\it Dipartimento di Fisica e Astronomia and INFN sezione di Catania}
\baselineskip=10pt
\centerline{\footnotesize\it Universita' di Catania, Catania I-95123, Italy}
\centerline{\footnotesize\it E-mail: Alessandro.Pluchino@ct.infn.it}
\vspace*{15pt}          
\centerline{\footnotesize ANDREA RAPISARDA}
\baselineskip=12pt
\centerline{\footnotesize\it Dipartimento di Fisica e Astronomia and INFN sezione di Catania}
\baselineskip=10pt
\centerline{\footnotesize\it Universita' di Catania, Catania I-95123, Italy}
\centerline{\footnotesize\it E-mail: Andrea.Rapisarda@ct.infn.it}
\vspace*{0.225truein}

\vspace*{0.25truein}

\abstracts{We study the continuum opinion dynamics of the compromise model
of Krause and Hegselmann for a community of mutually interacting agents,
by solving numerically a rate equation. The opinions are here represented
by bidimensional vectors with real-valued components.
We study the situation starting from a uniform
probability distribution for the opinion configuration and
for different shapes of the confidence range.
In all cases, we find that the thresholds for consensus and cluster merging
either coincide with their one-dimensional counterparts, or are very close to them.
The symmetry of the final opinion configuration, when more clusters survive,
is determined by the shape of the opinion space. If the latter
is a square, which is the case we consider, the clusters in general occupy the sites of a square lattice,
although we sometimes observe interesting deviations from this general pattern,
especially near the center of the opinion space.}{}{}

\vspace*{5pt}



\vspace*{1pt}\textlineskip      

\section{Introduction}         
\vspace*{-0.5pt}
\noindent

If two individuals are supporters of the same football team,
they may like to discuss about football; on the other hand, if two persons
have opposite political views, they will hardly discuss about politics.
Suppose now to have two individuals who are fans of the same football team but
support opposite political parties: will they talk to each other or not? Will
their affinity in football win over their contrasts in politics?
The question sounds ambiguous: some pairs of people will interact,
others will not. What we would like to stress here is the rather obvious fact that
the tendency of people to effectively communicate with each other is usually
favoured by affinities and hindered by contrasts. So far so good, this is
basic sociology. Indeed, Axelrod \cite{Axel} had this in mind when he proposed
his model for the dissemination of culture, to explain how different cultural islands could be generated from a local tendency to convergence. For Axelrod, "culture" is
modelized as an array of features, each feature being specified by "traits",
which are expressed by integers. The number of features or dimensions
is nothing but the number of components of a vector, and two persons interact
if and only if they share at least one common feature (i.e. the same value of
the corresponding vector component). In this model, two persons are the culturally closer
the more features they have in common, and the number of common features is related to the
probability of the two individuals to interact. Starting from the Axelrod model,
a number of simple agent-based models have been devised, mostly by physicists.
This should not be surprising, as these models are essentially cellular
automata, which physicists are used to play with.
Among these models we mention those of Sznajd \cite{Sznajd}, Deffuant et
al. \cite{Deff}, Krause and Hegselmann \cite{HK}, Wu and Huberman \cite{huberman},
Pluchino et al. \cite{pluchino}.
For a review of the recent research activity in this field we refer to
\cite{staufrev,extreme}.

In this way proper physical questions could be addressed, which concern not
just the behaviour of single individuals but the collective behaviour of a community.
In fact, if the behaviour of a person is of little relevance for quantitative
scientific investigations, and essentially unpredictable,
the global organization of many mutually interacting
subjects presents general patterns which go beyond specific
individual attributes and may emerge in several different contexts.
One can then hope that quantities like
averages and statistical distributions may characterize not just specific situations
but large classes of systems. This explains why in the last years
physicists tried to use their expertise to investigate social systems, where the elementary
interacting objects are the people (or "agents") and the "charge" is represented by their
opinion, although the latter is strictly speaking not a measurable quantity.

In all above-mentioned models opinions are modelized as numbers, integer or
real. One starts by assigning randomly a number to every agent of the system.
Then the dynamics starts to act, and the agents rearrange their opinion
variables, due to mutual discussions. At some stage, the system reaches
a configuration which is stable under the dynamics; this final configuration
may represent consensus, with all agents sharing the same opinion, polarization,
with two main clusters of opinions ("parties") or fragmentation, where
several opinion clusters survive. However, as we explained at the beginning,
a discussion between two persons is not simply stimulated by their common view/preference
about a specific issue, but it in general depends on the global affinity of the two
persons, which is influenced by several factors. So, for a more realistic modeling of
opinion dynamics, one should represent the opinions/attitudes like vectors, as Axelrod
did, and not like scalars. In a forthcoming paper \cite{sznajdmul},
K. Sznajd-Weron and J. Sznajd
assign two Ising-like spin variables to each agent, referring to
the opinion about a personal issue and an economic one.
In a very recent work on the
Deffuant model \cite{jacobmeier}, the opinion has several components
which are integers and the agents sit on the sites of a Barab\'asi-Albert
network. However, a systematic study of the problem of multidimensional opinion
dynamics is still missing, and the aim of this paper is to try to fill up this gap.

In this paper we deal precisely with this kind of problem; opinions are
bidimensional vectors $\vec{S}=(x,y)$, and the components can take any real value in some finite
interval (e.g. $[0:1]$). Instead of formulating a new opinion dynamics, it is for us
more important to check what happens if we use one of the existing
models. In this way one can compare the results for vector opinions with those
relative to standard scalar opinions. We chose to adopt the opinion dynamics
proposed by Krause and Hegselmann \cite{HK} (KH), which has recently been subject of
intense investigations \cite{san1,san2}. The KH model is based on bounded
confidence, i.e. on the presence of a parameter $\epsilon$, called confidence
bound, which expresses the compatibility of agents. If the opinions of two
agents $i$ and $j$ differ by less than $\epsilon$, their positions are close
enough to allow for a discussion between $i$ and $j$ which eventually leads to a
variation of their opinions, otherwise the two agents do not interact with each
other. A society, or community, is modelized as a graph, where the vertices
represent the agents and the edges relationships between agents. So we say that
two agents are friends and could eventually talk to each other if there is an
edge joining the two corresponding vertices (in graph language, if the two
vertices are neighbours).

The dynamics of the model is very simple: one chooses at random one of the
agents and checks how many neighbours of the selected agent are compatible with it.
Next, the agent takes the average opinion of its compatible neighbours.
The procedure is repeated by selecting at random another agent and so on.
The type of final configuration reached by the system depends on the value of
the confidence bound $\epsilon$. For a society where everybody talks to
everybody else, consensus is reached for $\epsilon>\epsilon_c$, where
$\epsilon_c\sim 0.2$ \cite{san2}. Whereas one may expect that the value of
the consensus threshold be strictly related to the type of graph adopted to
modelize society, it turns out that it can take only one of two possible values,
$\epsilon_c\sim 0.2$ and $1/2$,
depending on whether the average degree (= number of neighbours)
of the graph diverges or stays finite
when the number of vertices goes to infinity \cite{san2}.

Here we want to apply the dynamics of the KH model
to vector opinions, using a continuum opinion distribution and
integrating the corresponding rate equation.
We focus on a society where everybody talks to everybody else,
because in this case the evolution of the system can be described
by a simple rate equation, that one can easily solve numerically. This procedure has
already been used to investigate the compromise model of Deffuant et al. with
one-dimensional opinions \cite{bennaim}.
The advantages of the evolution equation over Monte Carlo simulations
are that one can (in principle) deal with a system with
arbitrarily many agents and can better resolve the crucial steps of the time
evolution of the system, especially when opinion clusters merge.
Furthermore, as we will see, the final cluster configurations obtained for a continuum distribution will be much more symmetric and regular.

\section{The model}

The opinion space is represented by the points $(x,y)$ of a bidimensional manifold, that for us is a square: $[0,1]\times [0,1]$.
The continuum distribution of the opinions among the agents is expressed by the
function $P(x,y,t)$, such that $P(x,y,t)dx\,dy$ represents the fraction of agents
whose opinions have components in the intervals $[x, x+dx]$ and $[y, y+dy]$.
The integral of the distribution $P(x,y,t)$ over the whole opinion space is of course one.
The dynamics of KH, as well as that of Deffuant et al., can simply be extended
to the multi-dimensional case. We must however be careful to the definition of
bounded confidence, and the corresponding parametrization. The crucial point is
the concept of "closeness" between agents. Shall one consider the two opinion
components independently or jointly? In the first case one
can assume that two agents are compatible if either the $x$- or the $y$-components of their
opinions are close enough to each other, even if the other components
are far apart.
However, in this case, the presence of several different issues would not
represent a change with respect to the standard situation with scalar opinions.
In fact, the issues can be considered separately and each opinion component
would evolve independently of the others, so that one would just have to compose
the results obtained for each single issue. As we said in the introduction, this
is not what we would like to have, and it is not what happens in society.
The closeness of the agents depends on the general affinity between them,
and the affinity has to do with all issues. So, two agents are compatible if
both their opinion components are close to each other.
The shape of the confidence range, i. e. of the set of points in the opinion space
which represent all opinions compatible with that of some agent, can be arbitrarily
chosen: we took the two simplest cases of the square and the circle.

We have then two possible scenarios:

\begin{enumerate}
\item{squared opinion space, squared confidence range;}
\item{squared opinion space, circular confidence range.}
\end{enumerate}

\begin{figure}[htb]
\vspace*{13pt}
\centerline{
\epsfig{file=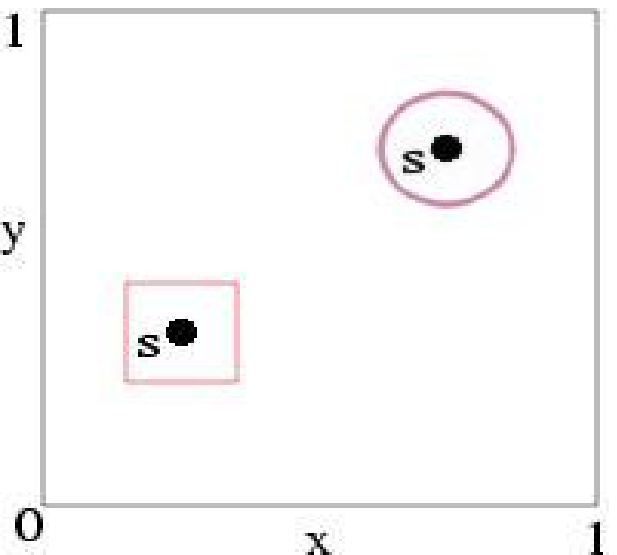,scale=0.5}
}
\vspace*{10pt}
\fcaption{\label{fig1}Simplest scenarios for our model: the opinion space is a square and the confidence range can be a square or a circle.}
\vspace*{13pt}
\end{figure}

The first scenario is illustrated in Fig. \ref{fig1}: the large square is the
opinion space, the black dot is the opinion $\bf{S}$ of an agent $A$ and the smaller square
centered at $\bf{S}$ is the (symmetric) confidence range of the agent. All agents
whose opinions lie within the square are compatible with $A$ and can interact
with it. For the second scenario, the square is replaced by a circle of radius
$\epsilon$.
We now come to the evolution equation of the system. We start by examining the
continuum version of the standard one-dimensional KH model, in order to test the goodness of our procedure for the
numerical solution of the rate equation,
by making comparisons with known results from Monte Carlo simulations.
After that, we turn to the bidimensional case.

\section{Results for scalar opinions}

The opinion space is the range $[0,1]$, the confidence bound $\epsilon$.
The opinion distribution among the agents is given by the function $P(x,t)$.
The rate equation of the KH model is

\begin{equation}
\frac{\partial}{\partial t}P(x,t)=\int_0^1 dx_1\,P(x_1,t)\Big[\delta\Big(x-
\frac{\int^{x_1+\epsilon}_{x_1-\epsilon}dx^{\prime}\,x^\prime\,P(x^\prime,t)}
{\int^{x_1+\epsilon}_{x_1-\epsilon}dx^\prime\,P(x^\prime,t)}\Big)-\delta(x-x_1) \Big]
\label{eq1}
\end{equation}

The two $\delta$'s in the equation represent the only two contributions to the variation
of $P(x,t)$ around the opinion $x$ in the time interval $[t,t+dt]$. In fact, take
all agents in the range $[x_1,x_1+dx_1]$, with $x_1$ in $[0,1]$. The
complicated ratio of integrals inside the first $\delta$ is nothing but the
average
opinion $\bar{x}$ of all agents whose opinions are compatible with $x_1$. As we explained
above, $\bar{x}$ is then precisely the new opinion of the agents. So, whenever
$\bar{x}=x$, there will be new agents with opinion in the range $[x,x+dx]$. On
the other hand, if $x_1=x$, the final opinion will be in general different from
$x$, so those agents will leave the range $[x,x+dx]$. The total balance of the
two contributions is expressed by the integration over all values of $x_1$, which
cover the opinion range $[0,1]$.

We immediately see from the equation that the dynamics conserves the
total population of agents, $N(t)=\int_0^1\,P(x,t)dx$, as it should be for the
physical interpretation. In fact, if we integrate both sides of Eq. (\ref{eq1})
over $x$ on the opinion range $[0,1]$, we obtain

\begin{equation}
\frac{\partial}{\partial t}\int_{0}^{1}dx\,P(x,t)=\int_{0}^{1}\int_0^1 dx\,dx_1\,P(x_1,t)
\Big[\delta\Big(x-\bar{x}(x_1,t) \Big)-\delta(x-x_1) \Big]
\label{eq10}
\end{equation}
where, for simplicity, we indicate with $\bar{x}(x_1,t)$ the ratio of integrals within
the first $\delta$.

If we perform the integral over the variable $x$ on the right hand side, we see
that the only dependence on $x$ is contained in the
two $\delta$'s; by integrating them in the range $[0,1]$ they
give equal contributions but opposite in sign ($1-1=0$).
The derivative with respect to time of the norm $N(t)$ (left-hand side) is then zero and the
norm keeps its initial value $N(0)$ during the whole evolution (we set $N(0)=1$).

It is not so straigthforward, though relatively simple, to check that also the
first moment of the opinion distribution, i. e. the average value of the opinion
of the system, is conserved by the dynamics. For this purpose one should
multiply both sides of Eq. (\ref{eq1}) by $x$ and integrate over $x$ in $[0,1]$,

\begin{equation}
\frac{\partial}{\partial t}\int_{0}^{1}dx\,x\,P(x,t)=\int_{0}^{1}\int_0^1 dx\,dx_1\,x\,P(x_1,t)
\Big[\delta\Big(x- \bar{x}(x_1,t) \Big)-\delta(x-x_1) \Big].
\label{eq11}
\end{equation}

Again, we first perform the integration over the variable
$x$ on the right-hand side, and we obtain

\begin{equation}
\frac{\partial}{\partial t}\int_{0}^{1}dx\,x\,P(x,t)=
\int_{0}^{1}\,dx_1\,P(x_1,t)\Big[\bar{x}(x_1,t)-x_1 \Big].
\label{eq12}
\end{equation}

We remark that the KH dynamics, as well as the Deffuant dynamics,
is symmetric with respect to the transformation $x\rightarrow\, 1-x$, i. e.
the two wings of the opinion range are perfectly equal, and remain equal during
the whole evolution of the system. Let us now focus on
the integral on the right-hand side of Eq. (\ref{eq12}): we will show that it
vanishes. We just need to
transform the integration variable
$x_1$ in $1-x_1$. We get

\begin{equation}
\int_{0}^{1}\,dx_1\,P(x_1,t)\Big[\bar{x}(x_1,t)-x_1 \Big]=
-\int_{1}^{0}\,dx_1\,P(1-x_1,t)\Big[\bar{x}(1-x_1,t)-(1-x_1) \Big].
\label{eq13}
\end{equation}

\noindent Due to the above-mentioned symmetry, we have

\begin{equation}
P(1-x_1,t)=P(x_1,t)
\label{eq14}
\end{equation}

\noindent and

\begin{equation}
\bar{x}(1-x_1,t)=1-\bar{x}(x_1,t),
\label{eq15}
\end{equation}

\noindent so we finally obtain
\begin{equation}
\int_{0}^{1}\,dx_1\,P(x_1,t)\Big[\bar{x}(x_1,t)-x_1 \Big]=
\int_{1}^{0}\,dx_1\,P(x_1,t)\Big[\bar{x}(x_1,t)-x_1) \Big],
\label{eq16}
\end{equation}
which means that both integrals must yield zero. The conservation of the
average opinion of the community, which also holds in the model of Deffuant
\cite{bennaim}, has several important consequences: for instance, if we
start from a uniform opinion distribution, $P(x,t=0)={\rm const}$, for which
the average opinion is $1/2$, an eventual consensus can only be reached at the
central opinion $1/2$. Moreover, as the flat distribution keeps the symmetry
with respect to the central opinion $1/2$, the final configuration of the system
will be characterized by a symmetric pattern of opinion clusters to the right and the
left of $1/2$; if the number of clusters is odd, there will necessarily be
a cluster sitting exactly at $x=1/2$.

Let us now come back to the rate equation (\ref{eq1}).
In order to test numerically the behavior of the one dimensional KH model in the limit
of a continuum distribution of opinions $P(x,t)$, we have integrated the rate equation
using a standard fourth order Runge-Kutta algorithm. The opinion range $[0,1]$ has been discretized
in $1000$ bins and the accuracy in $P(x,t)$ was of $10^{-9}$. In all simulations
we started from a flat distribution, $P(x,t=0)={\rm const}$, as one usually does for these simple
consensus models. In this way, we assume that all opinions
are equiprobable, as it could well be in a community before people begin to talk
to each other.
The dynamics runs until the distribution $P(x,t)$ reaches a stationary state
(with an accuracy of $10^{-5}$) for a given value of the confidence bound.

\begin{figure}[hbt]
\vspace*{13pt}
\centerline{\epsfig{file=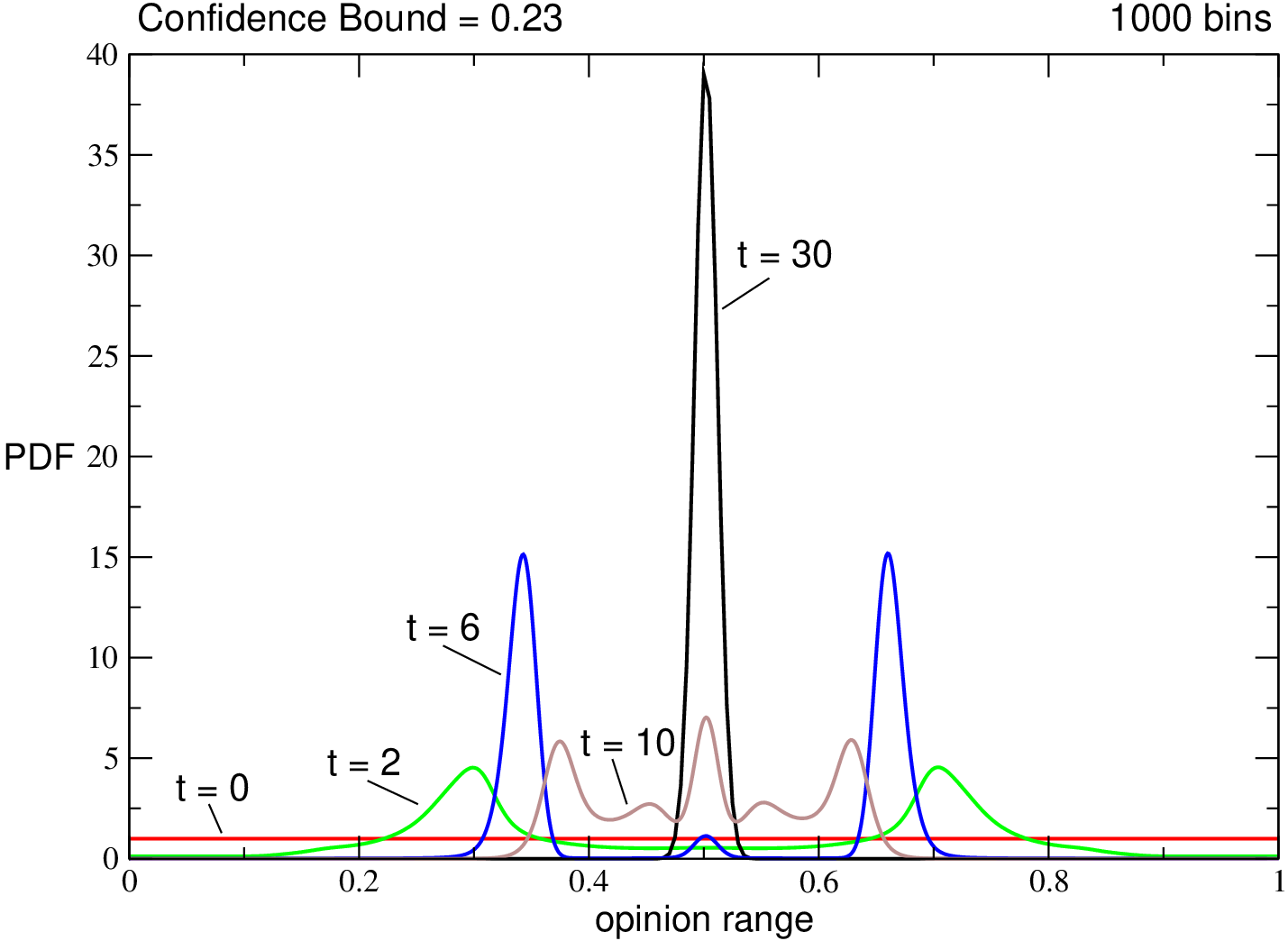, scale=0.4}}
\fcaption{\label{fig2}Time evolution of the opinion distribution for
 $\epsilon=0.23$. Initially the opinions condense in a two-clusters structure,
 then the two clusters slowly approach each other and merge, due to the presence of few
 agents with opinions lying near the center.
 After a long time, all agents end up in a single big cluster centered
 at $1/2$ (consensus).}
\vspace*{13pt}
\end{figure}

Fig. \ref{fig2} shows the time evolution of the opinion distribution, for $\epsilon=0.23$.
One clearly sees how the inhomogeneities at the two edges of the opinion range
create a perturbation which determines variations of the density profile, with
initial peaks forming close to the edges (not reported in the plot) and others which successively form towards the middle.
This is not a peculiar feature of the KH model; the time evolution
of the opinion dynamics of Deffuant et al. reveals the same pattern \cite{bennaim}.
For the chosen value of the confidence bound, two main peaks form rather
quickly. However
the configuration is not stable, and after a longer time the two peaks fuse in a
big cluster centered at the central opinion $1/2$. By looking at Fig. \ref{fig2},
we see that initially the two peaks are separated by a distance which exceeds the
confidence bound, so the corresponding agents are incompatible with each other
and should not interact. Why is then the two-peak configuration unstable in
this case? The reason has to do with the features of KH dynamics and it can be
better explained if we speak of single agents instead of opinion distributions.
In the KH model one agent is updated at a time; as we start from a flat opinion
distribution, there will be, at least initially, agents with opinions lying near
the central opinion $1/2$. Little by little these agents will accept the opinion
of the cluster which lies closest to it. But suppose that one agent $A$
lies between two different clusters of agents, $C_1$ and $C_2$, centered at
the opinions $s_1$ and $s_2$, respectively; suppose as well that $\epsilon$ is large
enough for $A$ to interact with the agents of $C_1$ and $C_2$, but
smaller than $|s_1-s_2|$. In this case, when we come
to update the opinion of the agent $A$, the latter will take the average
of the opinions of all agents in $C_1$ and $C_2$, which lies between $s_1$
and $s_2$.
On the other hand, when we update an agent $B$ in $C_1$, it
will be of course compatible with all other agents in $C_1$, but also with $A$
though not with $C_2$.
So, when one calculates the average of the opinions compatible with that
of $B$, it will not be exactly $s_1$, but it will depart
from $s_1$ by a tiny amount towards $s_A$. The same happens for the agents of
$C_2$ too. In conclusion, agent $A$ will keep lying between the two large
clusters, and the latter will slightly move towards each other due to the
intermediation of $A$. At some stage, the distance
of the two clusters will become smaller than the confidence bound $\epsilon$,
and all agents of the system will interact with each other, so they will all
take the average opinion $1/2$ of the whole system.
The process we have described shows that the KH dynamics spontaneously creates
a sort of hierarchy among agents that
otherwise behave in a perfectly identical way. Agent $A$ could be a political leader
which brings two parties to a mutual agreement.
As we have seen, the large clusters move very slowly;
in the limit of infinite agents, the presence
of a finite number of intermediary agents will lead a pair of clusters to merge
only after an infinitely long time. This happens not only at
the consensus threshold, but for all values of $\epsilon$ for which pairs of
clusters fuse according to the mechanism we have described, like the transition
from three to two final opinion clusters, from four to three, etc.

\begin{figure}[hbt]
\vspace*{13pt}
\centerline{\epsfig{file=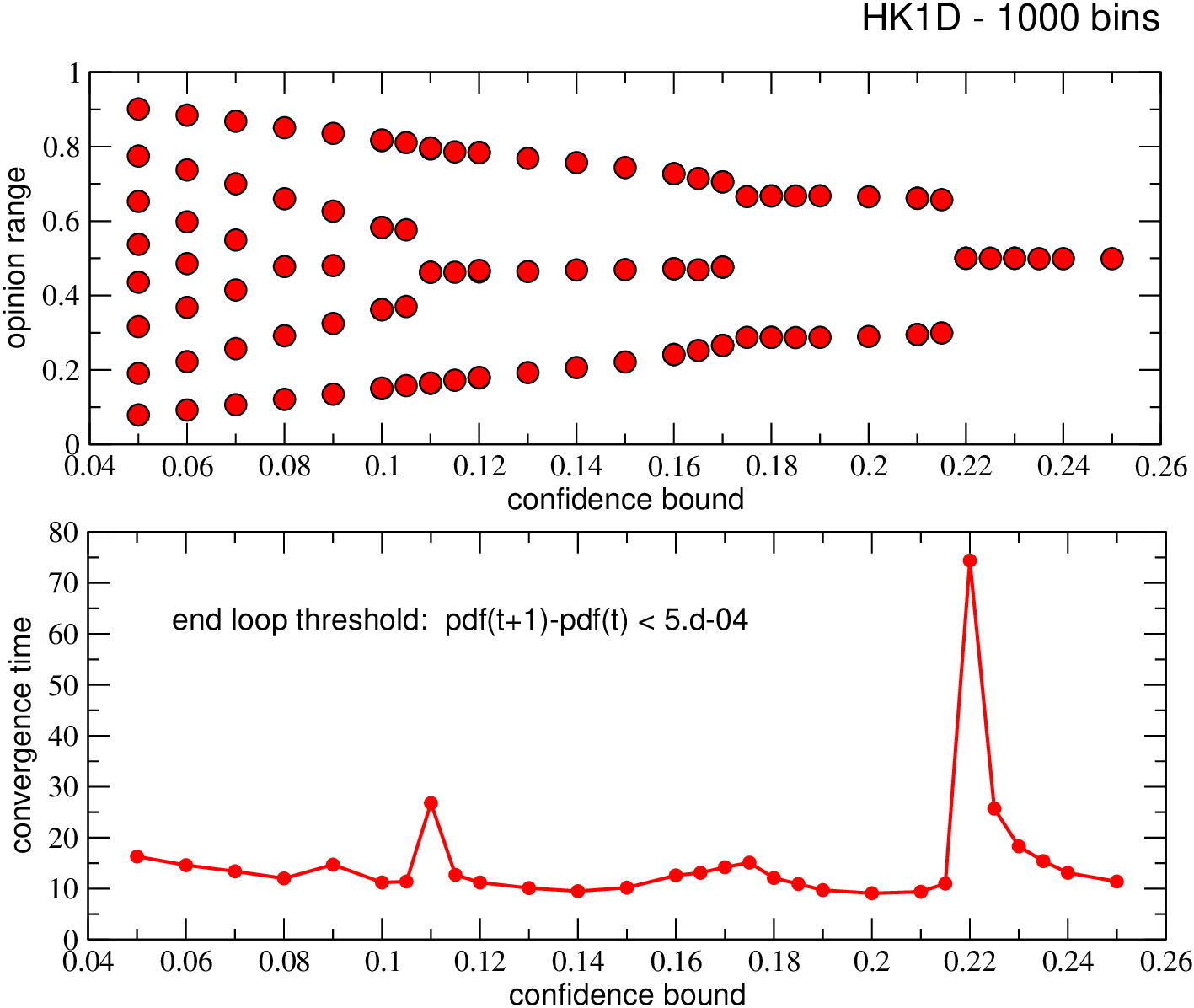, scale=0.4}}
\fcaption{\label{fig4}(Top) Final configuration of the system as a function of
$\epsilon$. The circles indicate the positions of the surviving clusters in
opinion space. (Bottom) Variation of the convergence times with $\epsilon$.
The narrow peaks are located in proximity of the
thresholds corresponding to cluster merging, at which the convergence time
diverges.}
\vspace*{13pt}
\end{figure}

In the model of Deffuant et al., where any agent interacts at a time with only one of its
compatible agents, not with all of them, there cannot be intermediary agents.
This implies that clusters can never interact with each other if their
separation in opinion space exceeds the confidence bound. This is actually the
reason why the consensus threshold in the model of KH is much lower than in
Deffuant\footnote{The consensus threshold for Deffuant, in a society where
everybody speaks with everybody else, is $\epsilon_c\sim 0.27$, in the sense
that for $\epsilon>\epsilon_c$
most agents belong to one large cluster and the others are either isolated or
form tiny groups. For $\epsilon>1/2$, instead, all agents end up in the same
cluster, also for other graph topologies \cite{unidef}.}.

Fig. \ref{fig4} (top) shows the position of the final clusters in opinion space
as a function of $\epsilon$. The threshold for consensus and that for the
transition from three to two final clusters are consistent with those determined
by means of Monte Carlo simulations. Notice the correspondence of the thresholds for cluster merging with the peaks in the convergence time (bottom).

\section{Results for vector opinions}

Now that we have tested the reliability of the numerical solution of the rate
equation,
we can proceed with the multidimensional case. The generalization of
Eq. (\ref{eq1}) is straightforward. For the sake of compactness, we will use the
following vector notation. The opinion vector {\bf S} is represented as
$\vec{x}$, an $n$-dimensional vector with components $x_1$, $x_2$, ... , $x_n$,
describing all points of the hypercube $[0,1]^n$. The opinion distribution can be
written as $P(\vec{x},t)$. In this way the rate equation can be immediately written

\begin{equation}
\frac{\partial}{\partial t}P(\vec{x},t)=\int_0^1 d\vec{x}_1\,P(\vec{x}_1,t)
\Big[\delta\Big(\vec{x}-
\frac{\int_{\Omega(x_1)}d\vec{x}_{0}\,\vec{x}_0\,
P(\vec{x}_0,t)}
{\int_{\Omega(x_1)}d\vec{x}_0\,P(\vec{x}_0,t)}\Big)-
\delta(\vec{x}-\vec{x}_1) \Big]
\label{eq2}
\end{equation}

Beware of the meaning of the symbols: the integral on the right-hand side
is a multiple $n$-dimensional integral and $\delta(\vec{x})=\delta(x_1)\,\delta(x_2)\,
...\delta(x_n)$. In the first multidimensional $\delta$, to get
the $i$-th component of the ratio of integrals one proceeds as follows.
The integrals must be both calculated within the hypervolume $\Omega(x_1)$, which
for us can be a hyperbox/hypersphere
centered at $\vec{x}_1$ and with side/diameter $2\epsilon$.
For the $i$-th component of the ratio, one has to replace the
term $\vec{x}_0$ inside the integral at the numerator with the corresponding
$i$-th component $x_{0i}$. It takes but a little work to show that the dynamics
of Eq. (\ref{eq2}), analogously as we have seen for scalar opinions, conserves
the zeroth and first moments of the opinion distribution.

Eq. (\ref{eq2}) is valid for an arbitrary number of dimensions.
We will deal here only with the bidimensional case, essentially for two reasons:

\begin{itemize}
\item{It is computationally cheap; if we discretize each opinion component in
   $N$ intervals, the hypercube becomes a grid with $N^n$ points, and
   as the number of operations is proportional to the number of points,
   the procedure becomes considerably slower for higher dimensions.}
\item{It is easy to present the results in figures, with three-dimensional or contour plots.}
\end{itemize}

\subsection{Squared confidence range}






Let us start with a squared opinion range.
We solved numerically Eq. (\ref{eq2}) for a few values of the confidence bound $\epsilon$,
here representing the half side of the square. As in the one-dimensional case, also this time
a fourth order Runge-Kutta integrator has been used and the simulations started from a flat ditribution,
$P(x,y,t=0)={\rm const}$. The squared ($x,y$) opinion space has been reduced to a grid of $100 \times 100$ bins ($200 \times 200$
bins have been also used in some simulations, in order to better estimate the consensus threshold) and
the accuracy in $P(x,y,t)$ was of $10^{-9}$. Even in this case, the dynamics runs until the distribution
$P(x,y,t)$ reaches a stationary state for a given value of the confidence bound.

\begin{figure}[htb]
\centerline{
\epsfig{file=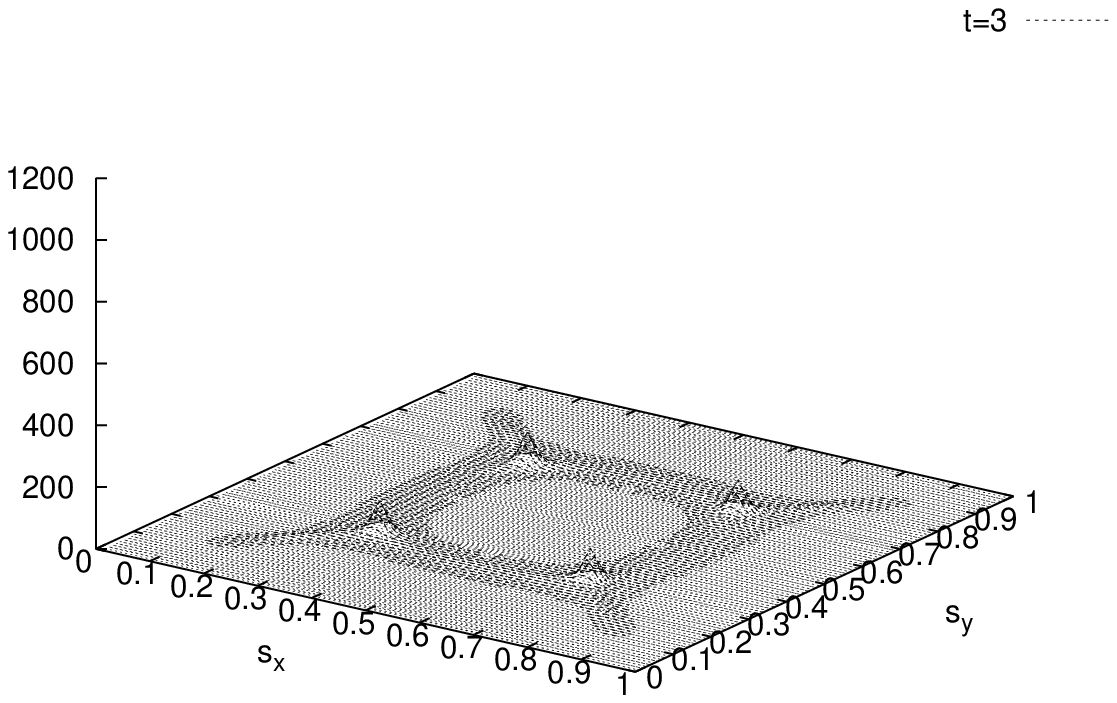, scale=0.47}
\epsfig{file=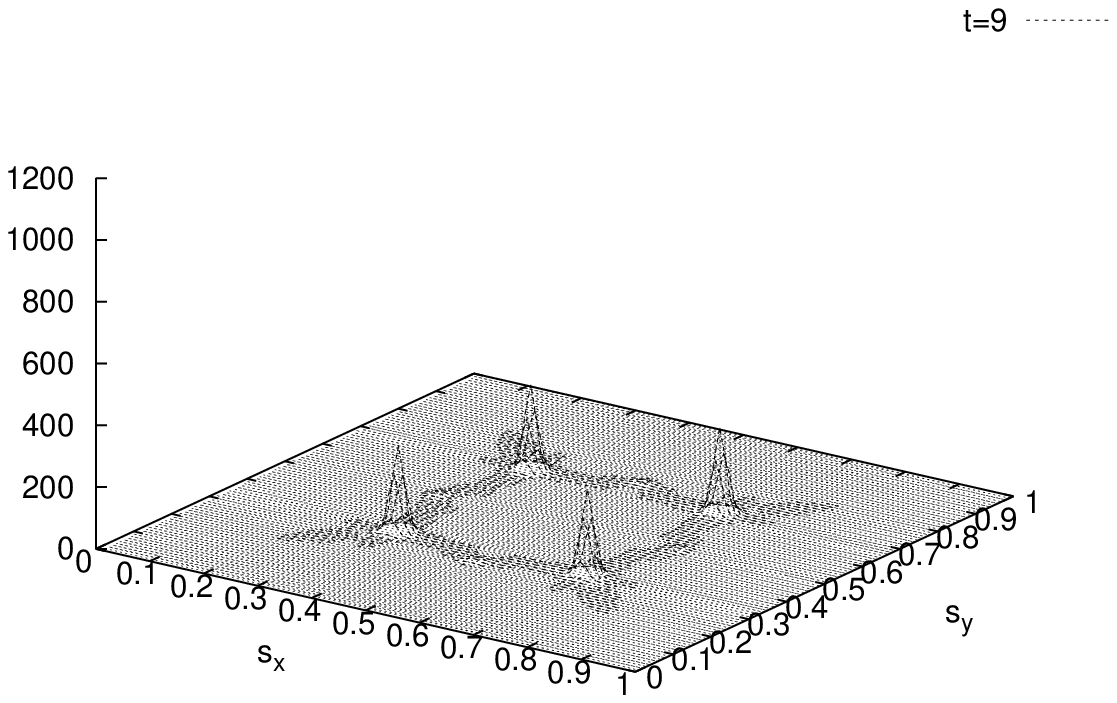, scale=0.47}
}
\centerline{
\epsfig{file=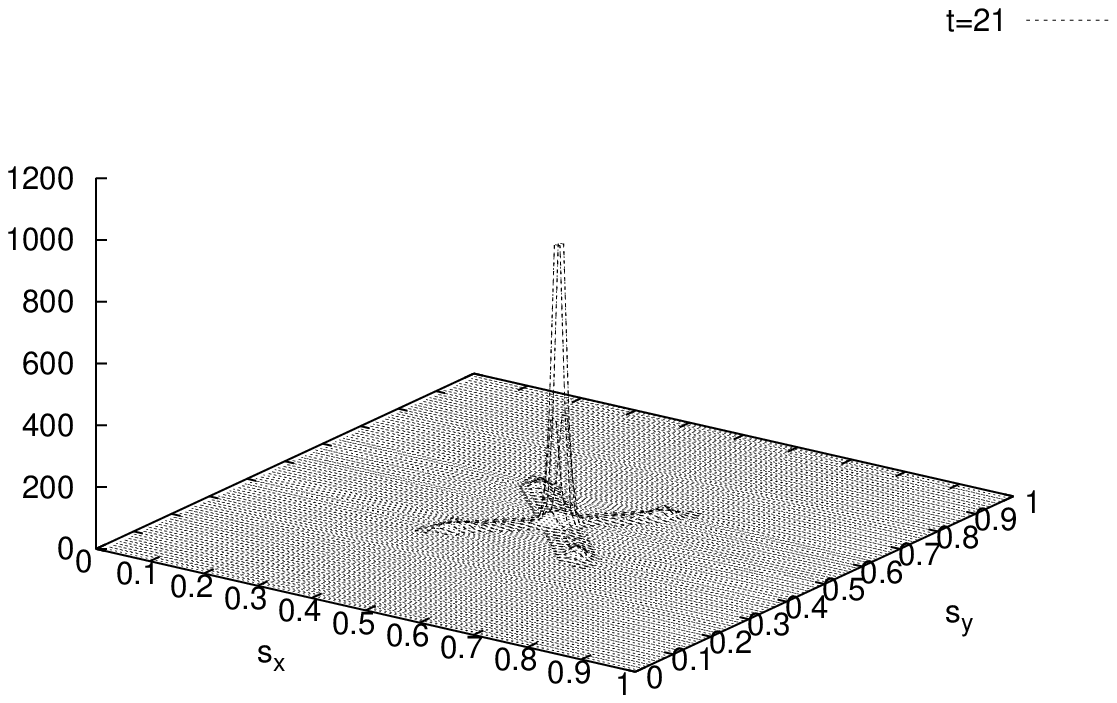, scale=0.47}
\epsfig{file=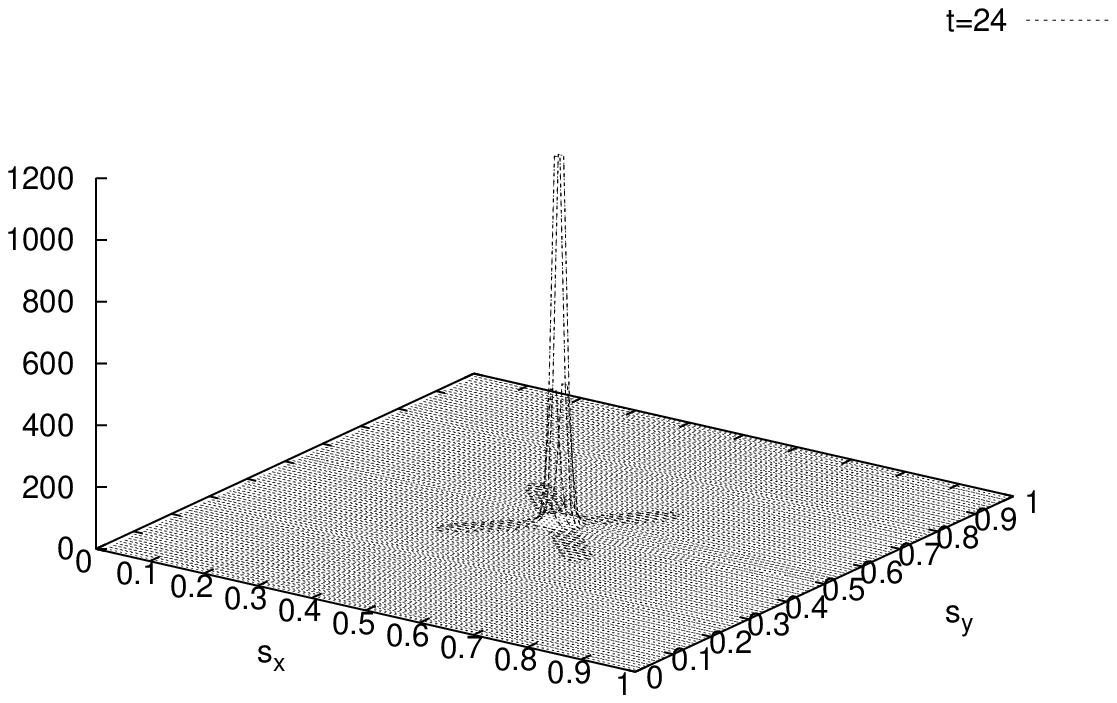, scale=0.47}
}
\fcaption{\label{fig5} KH dynamics with bidimensional real opinions and squared confidence range (from top left
to bottom right: t=3,9,21,24). The initial opinion distribution is uniform and the confidence bound $\epsilon=0.22$.}
\end{figure}
\begin{figure}[htb]
\centerline{
\epsfig{file=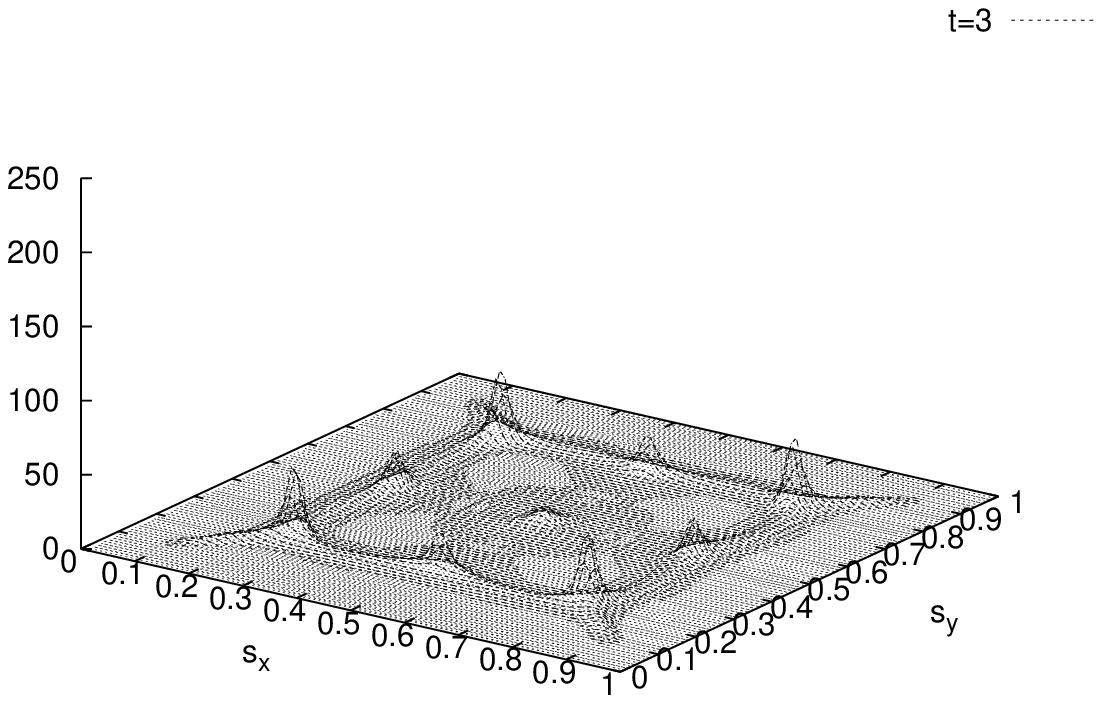, scale=0.47}
\epsfig{file=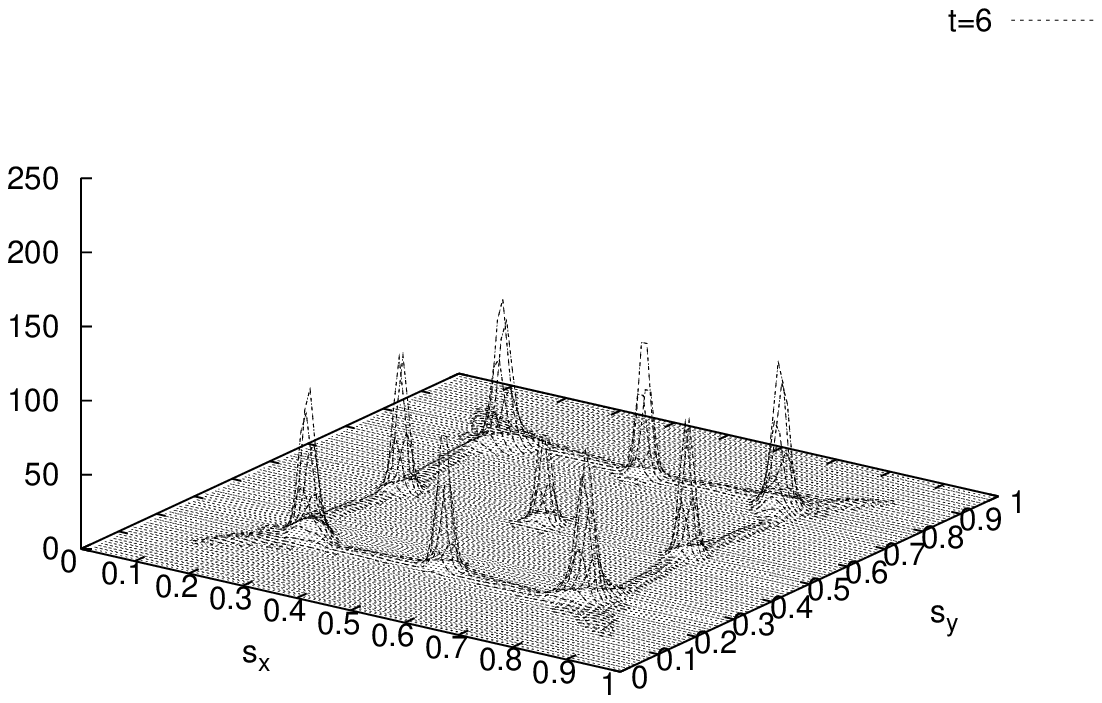, scale=0.47}
}
\centerline{
\epsfig{file=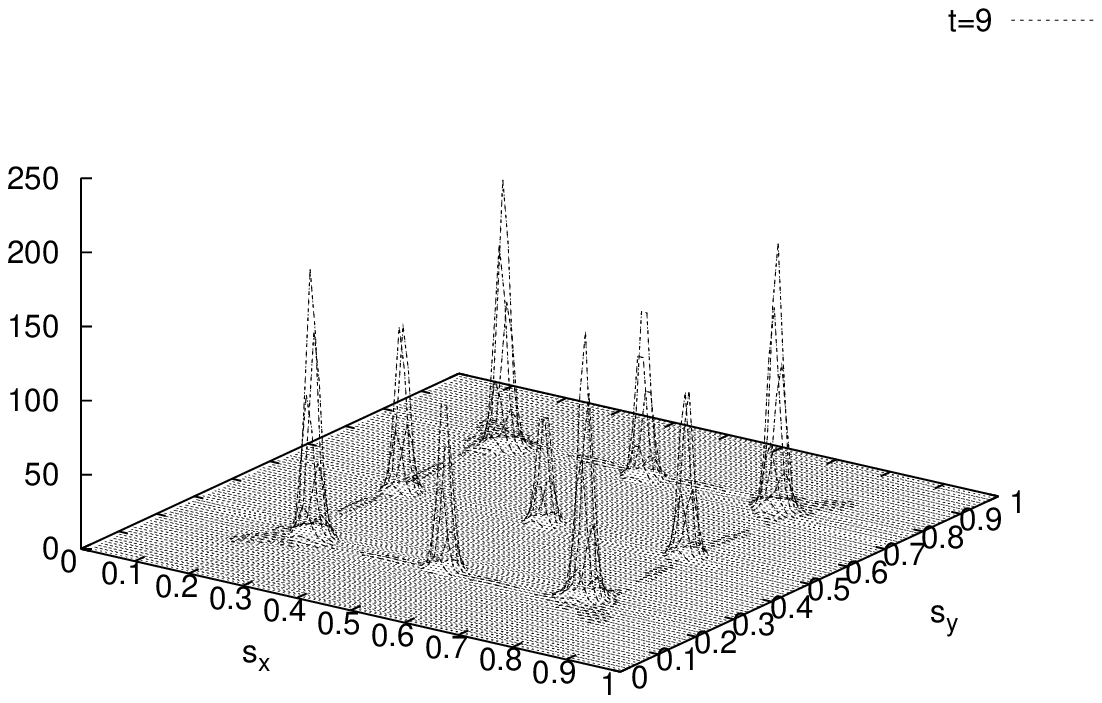, scale=0.47}
\epsfig{file=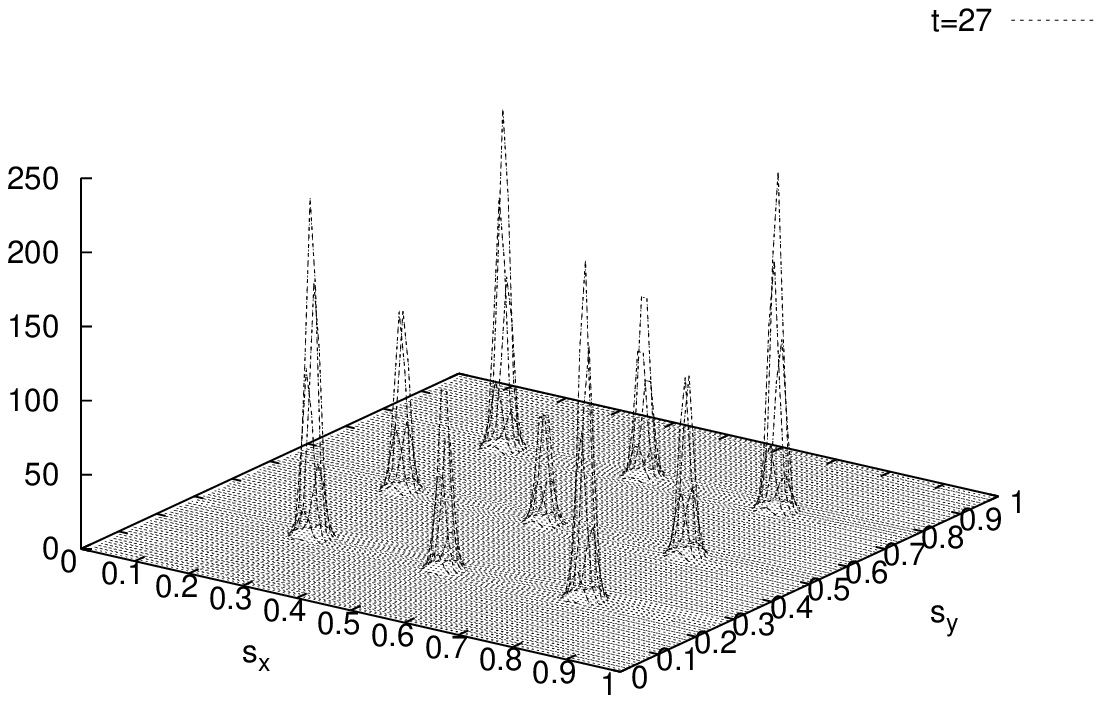, scale=0.47}
}
\fcaption{\label{fig6} As Fig. \ref{fig5} but for $\epsilon=0.15$ and t=3,6,9,27.}
\end{figure}
\clearpage
Fig. \ref{fig5} shows how an initially flat probability distribution evolves with time, for $\epsilon=0.22$.
We see that four major clusters are
 formed quite early, but after a sufficiently long time they fuse to a single central cluster.
The reason of such long-lived unstable states is again the fact that
clusters can interact with each other through intermediary
agents, as we have seen for scalar opinions.
If the confidence bound is small, we expect that many clusters survive. Fig. \ref{fig6}
shows that this is indeed the case. We notice the regular structure of the final
configuration; both the opinion space and the confidence range are squares, and
this symmetry is reflected in the opinion configurations, where the clusters sit on the sites of a square lattice.
From the figure one can see that the masses of the clusters are not equal. The four clusters near the vertices of the 
opinion space are the largest, followed by the four clusters near the centers of the edges of the square; 
the central cluster is the smallest. 

We have also tried to estimate the value of the consensus threshold $\epsilon_c$. Our
result, $\epsilon_c=0.215$, is consistent with the corresponding value for
standard one-dimensional opinions (the consistency refers to the
estimates determined through the rate equation; Monte Carlo simulations
deliver values closer to $0.2$ \cite{san2}).

\begin{figure}[hbt]
\vspace*{13pt}
\centerline{
\epsfig{file=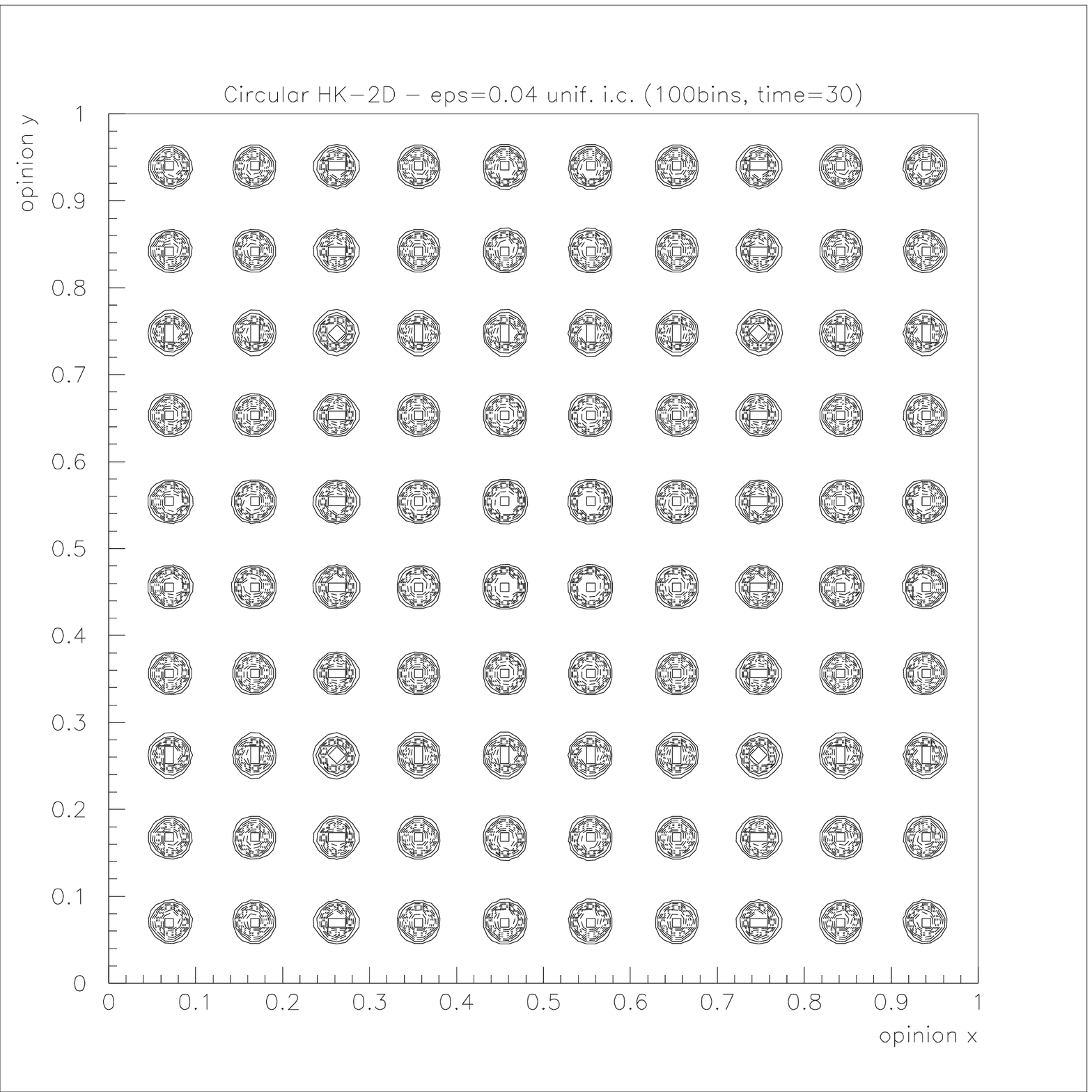, scale=0.25}
\epsfig{file=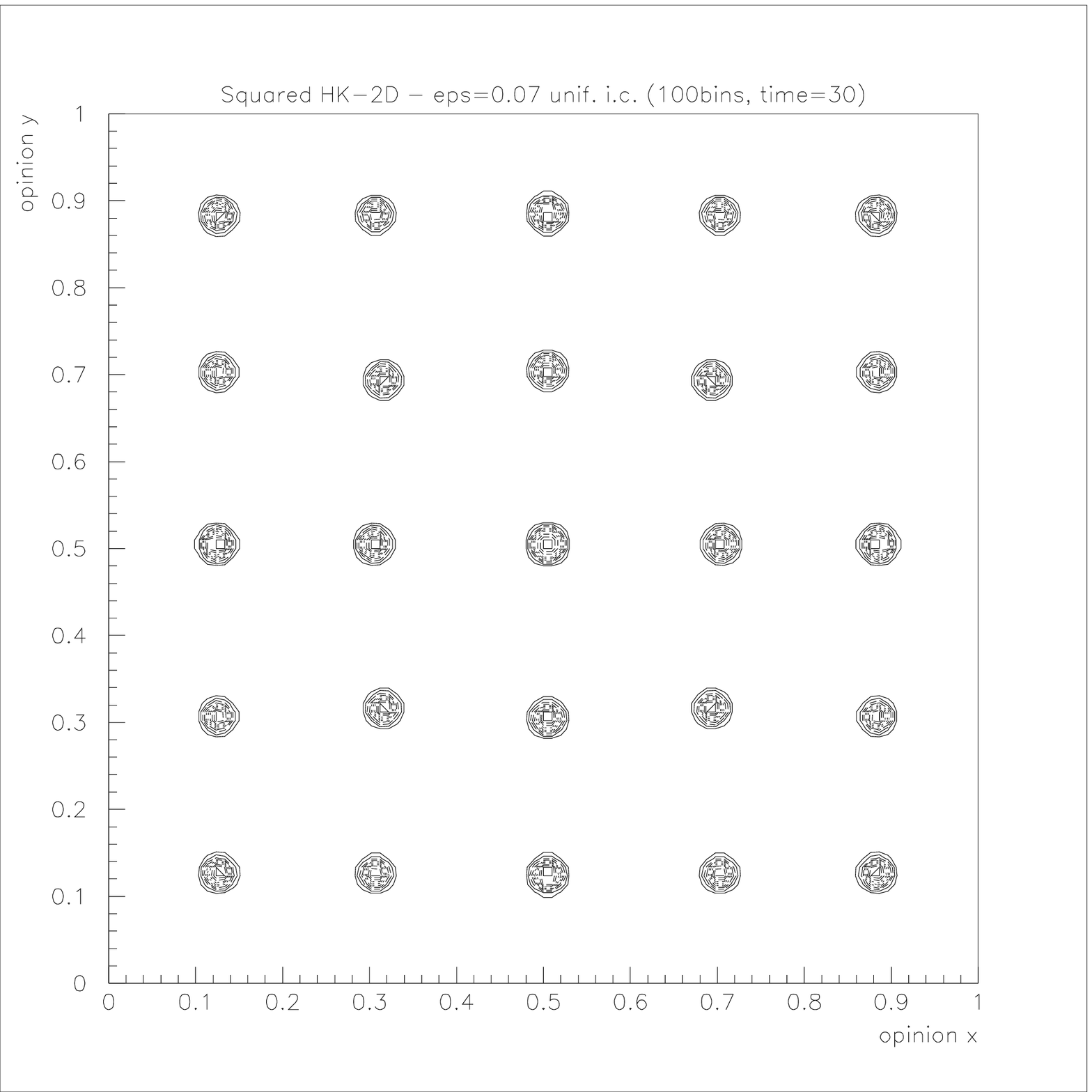, scale=0.25}
\epsfig{file=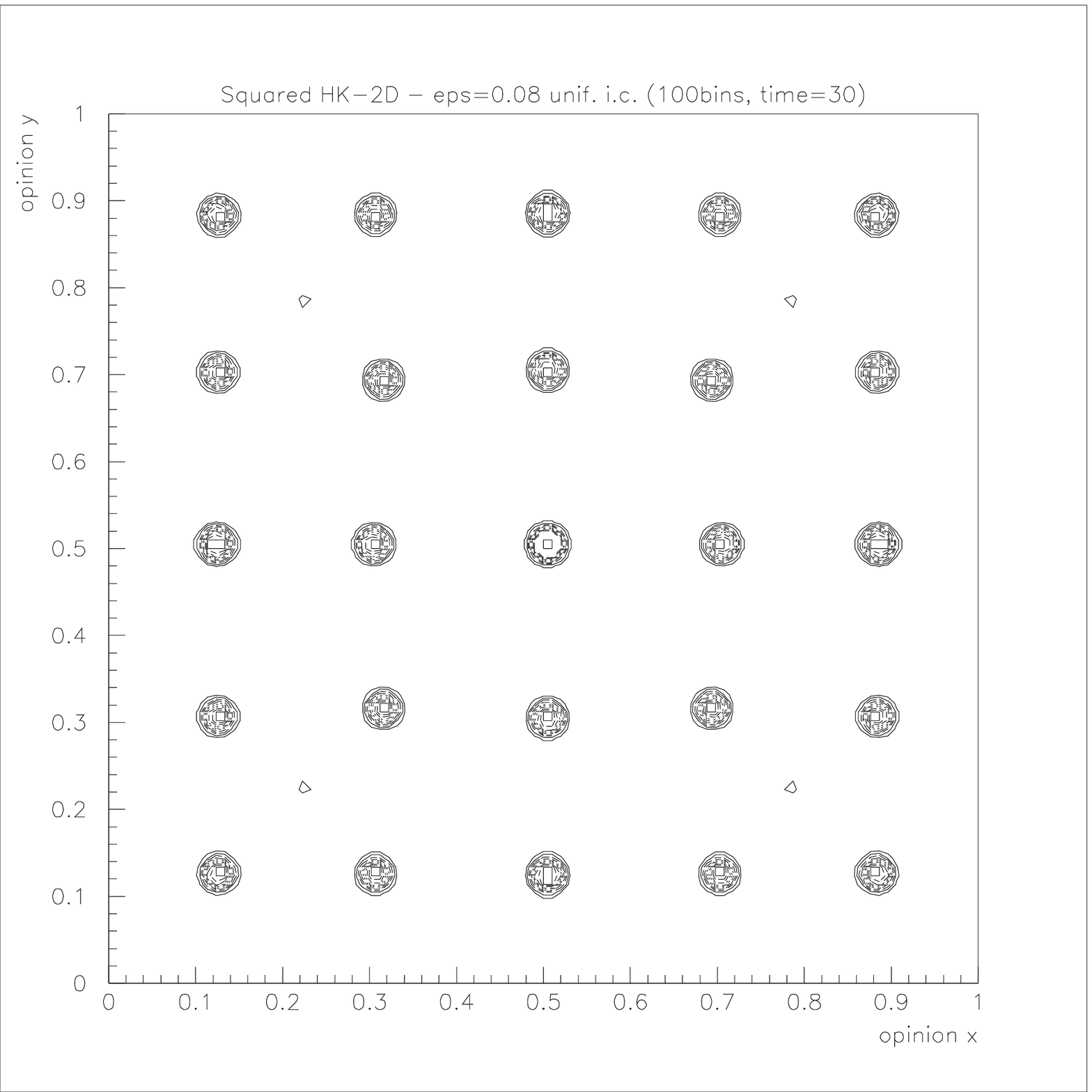, scale=0.25}
}
\centerline{
\epsfig{file=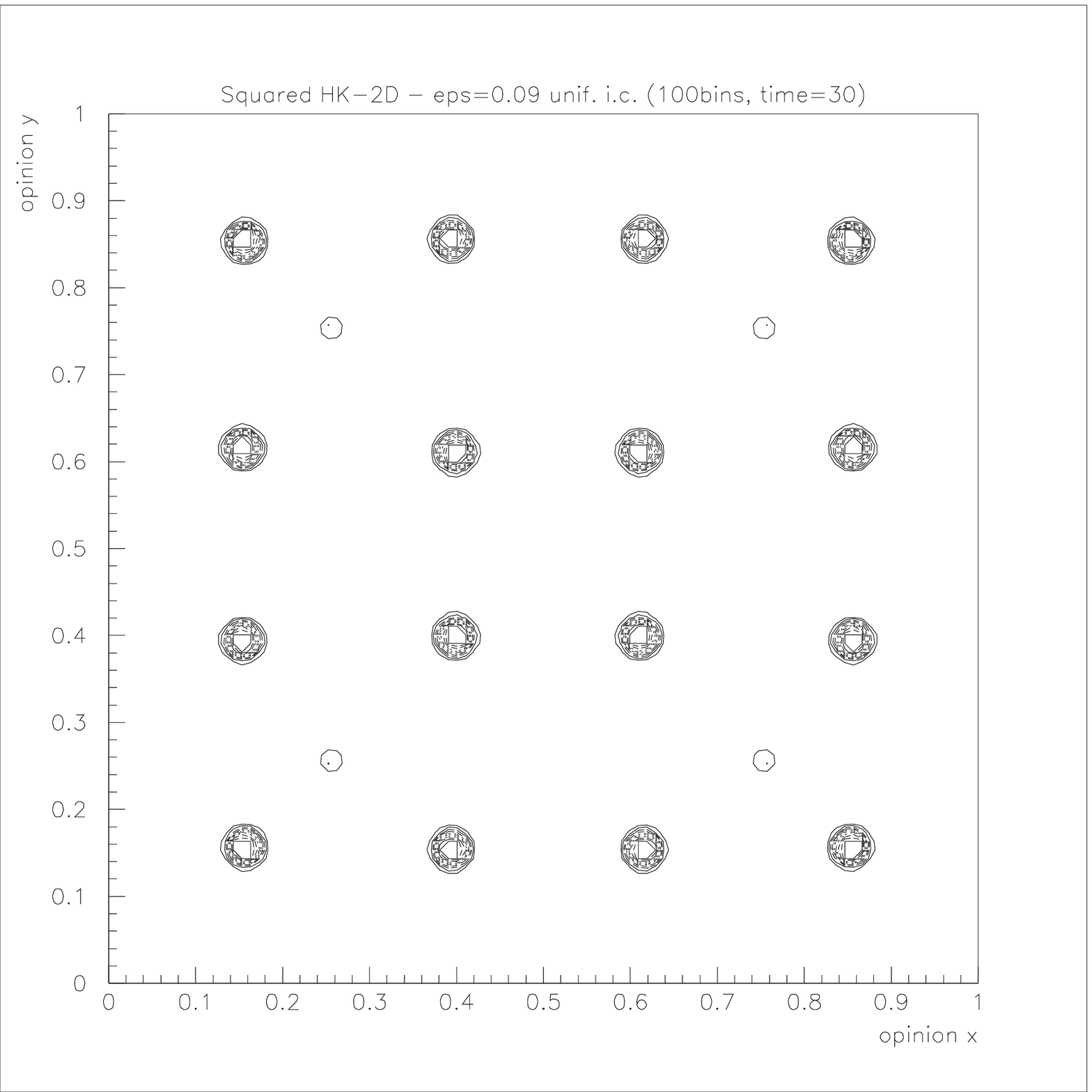, scale=0.25}
\epsfig{file=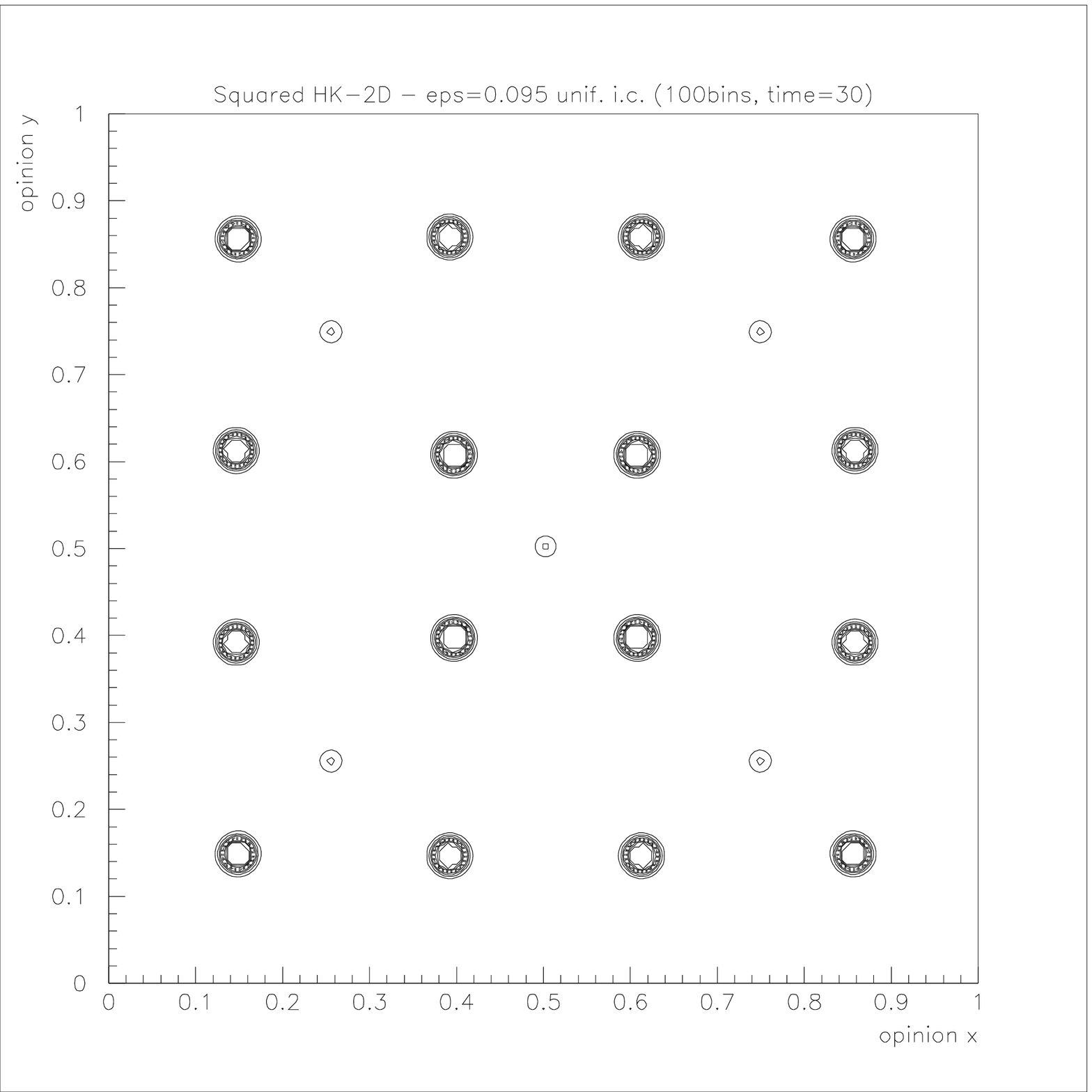, scale=0.25}
\epsfig{file=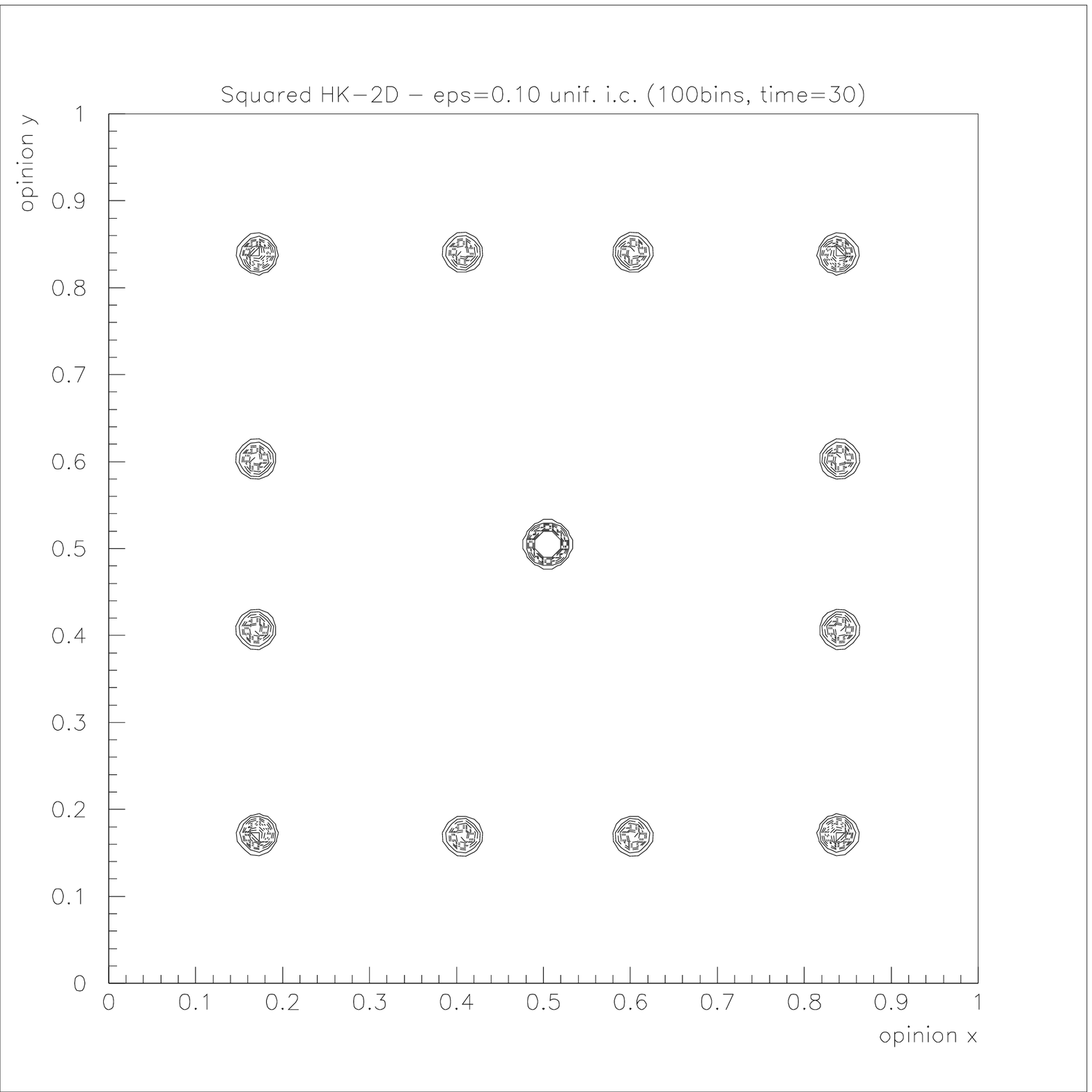, scale=0.25}
}
\centerline{
\epsfig{file=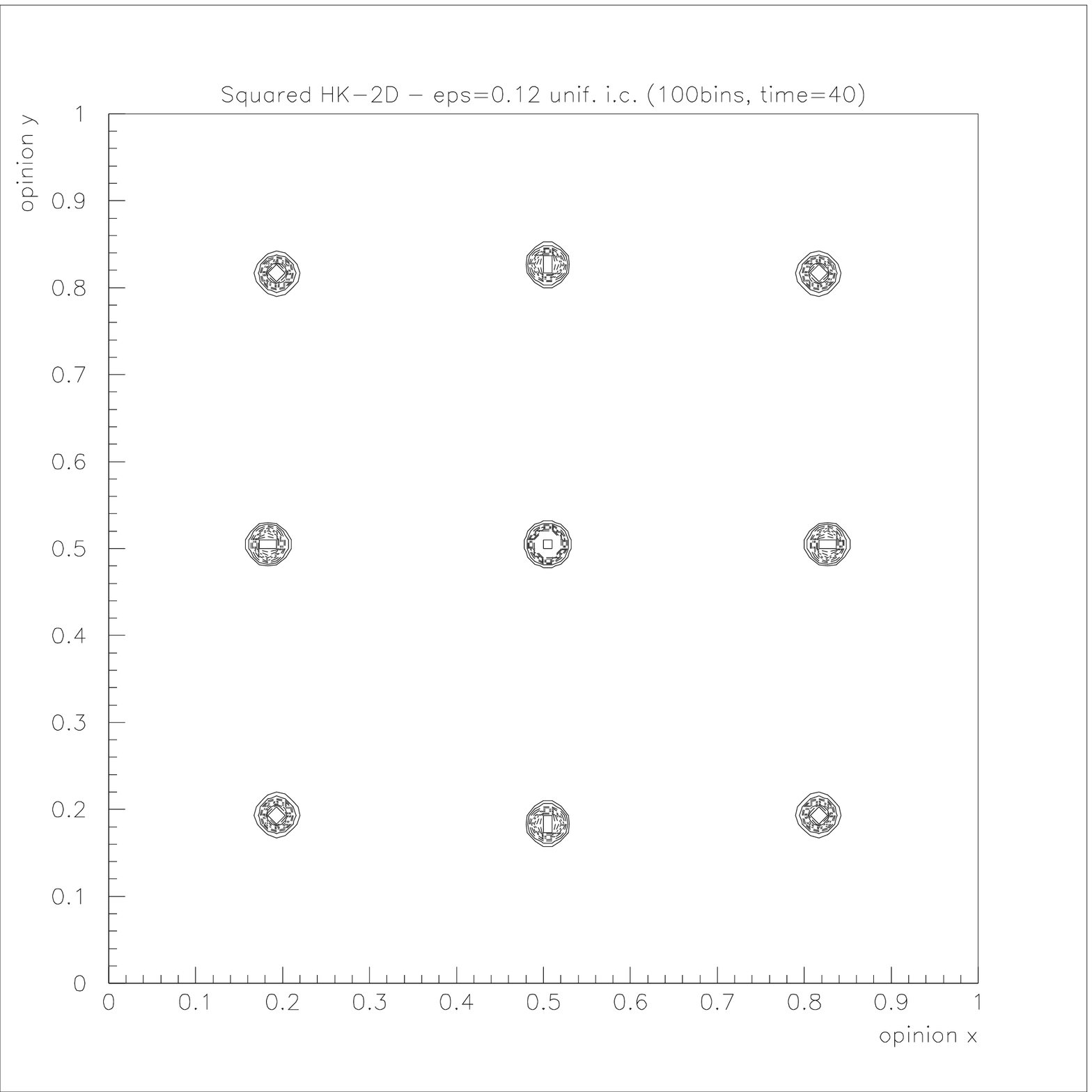, scale=0.25}
\epsfig{file=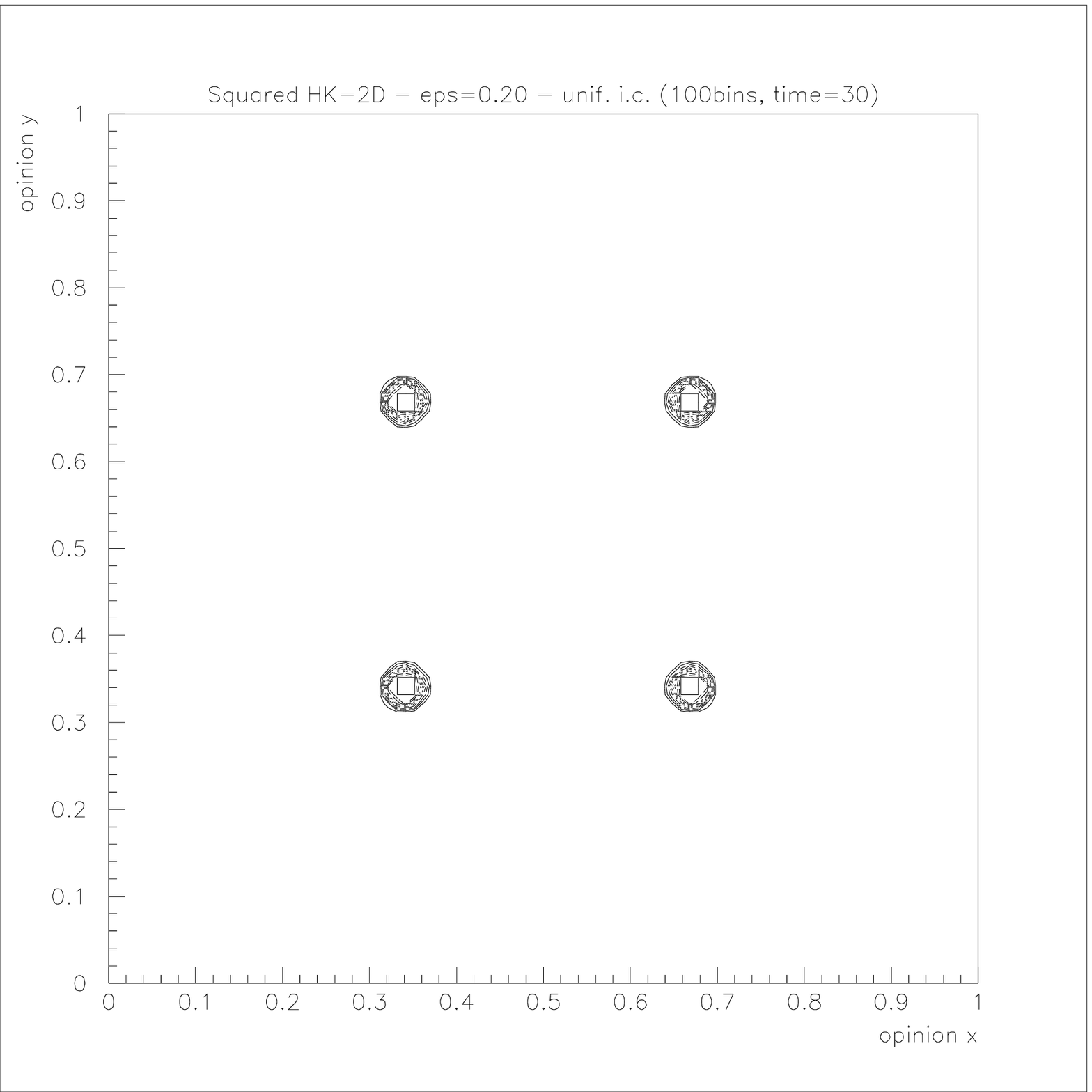, scale=0.25}
\epsfig{file=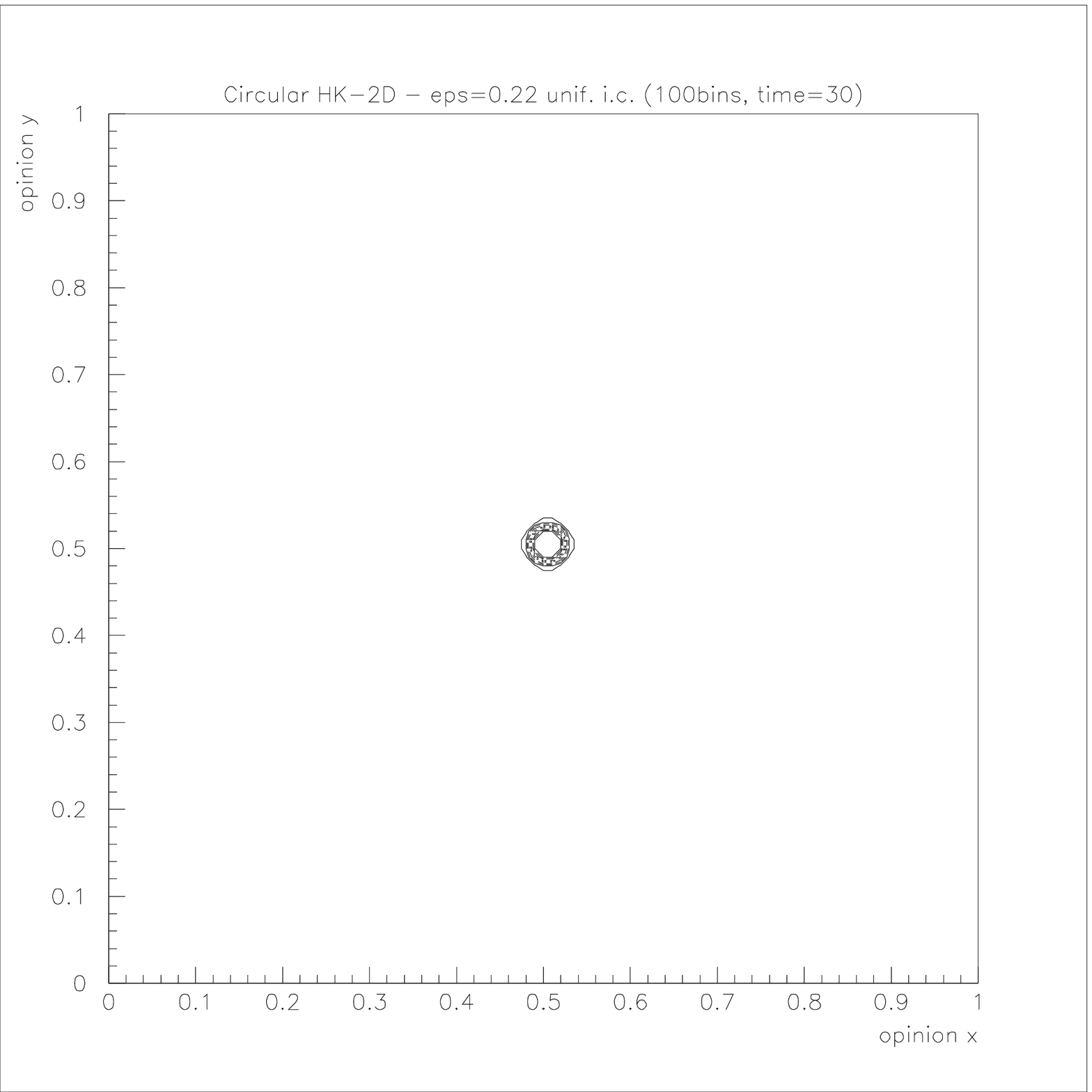, scale=0.25}
}
\vspace*{0.5cm}
\fcaption{\label{figsq} Final configurations of the KH model with bidimensional opinions and squared confidence range. From top left to bottom right: $\epsilon$=0.04,0.07,0.08; 0.09, 0.095,0.10; 0.12, 0.20,0.22.}
\vspace*{13pt}
\end{figure}

Fig. \ref{figsq} shows the final opinion configurations of the system
for several values of $\epsilon$. The figures are contour plots of the opinion distribution after many iterations.
We notice the symmetry of the configurations: the clusters sit on the sites of a square lattice.
We also find interesting variations of this scheme, however. As a matter of fact, we remark that in some cases also
small clusters survive, which lie on the sites of the dual lattice. In particular, when a small cluster
lies exactly on the center, it is likely to act as intermediate of the four large clusters which lie closest
to it, so that in the long run they fuse in a large central cluster, which explicitly breaks the lattice symmetry
of the configuration (as for $\epsilon=0.10$, for instance).
Such an anomalous feature can be viewed as an example of partial consensus below the critical treshold.

We have compared the final configuration of each opinion component
with that of the one-dimensional dynamics corresponding to the same value of $\epsilon$, for several values of $\epsilon$.
Most comparisons we have performed show a good
similarity, with just a few exceptions, which means that
the single opinion components - for a squared opinion range - evolve almost independently of each other.
The bidimensional dynamics is then, in most cases, effectively one-dimensional.

\clearpage

\subsection{Circular confidence range}

Let us now examine the situation when the confidence range is a circle
of radius $\epsilon$. In this case the two components $x,y$ are
necessarily correlated, thus we would expect
appreciable changes, both in the values of the thresholds for cluster merging and
in the symmetry of the final opinion configurations.

\begin{figure}[hbt]
\vspace*{13pt}
\centerline{
\epsfig{file=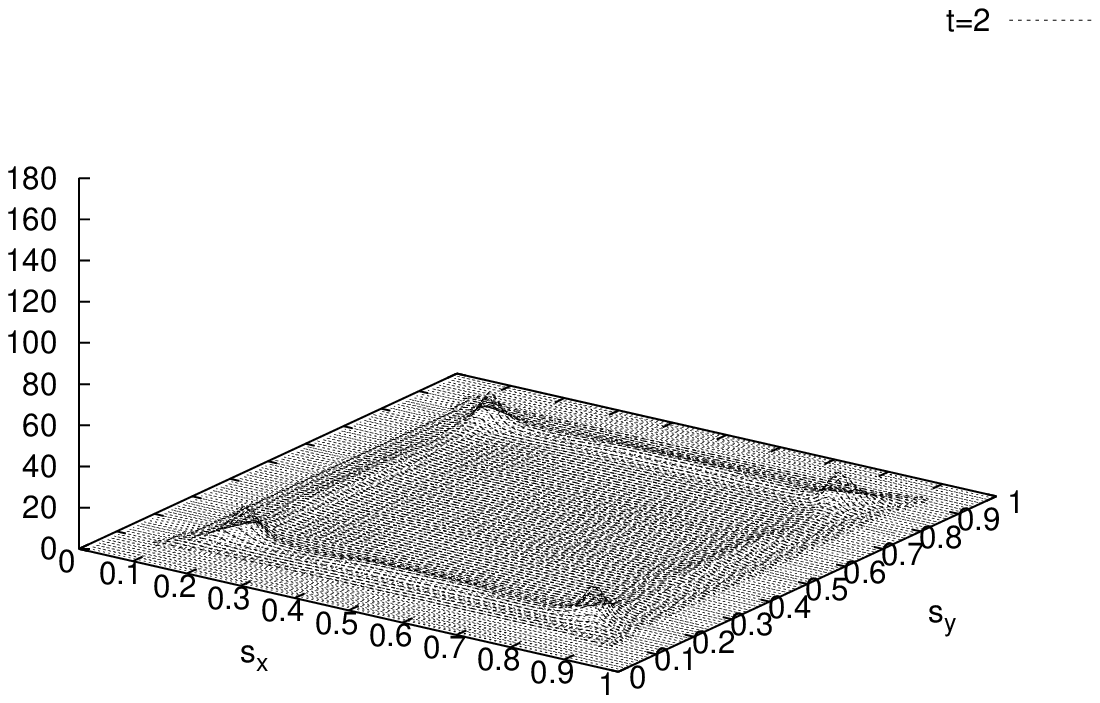, scale=0.47}
\epsfig{file=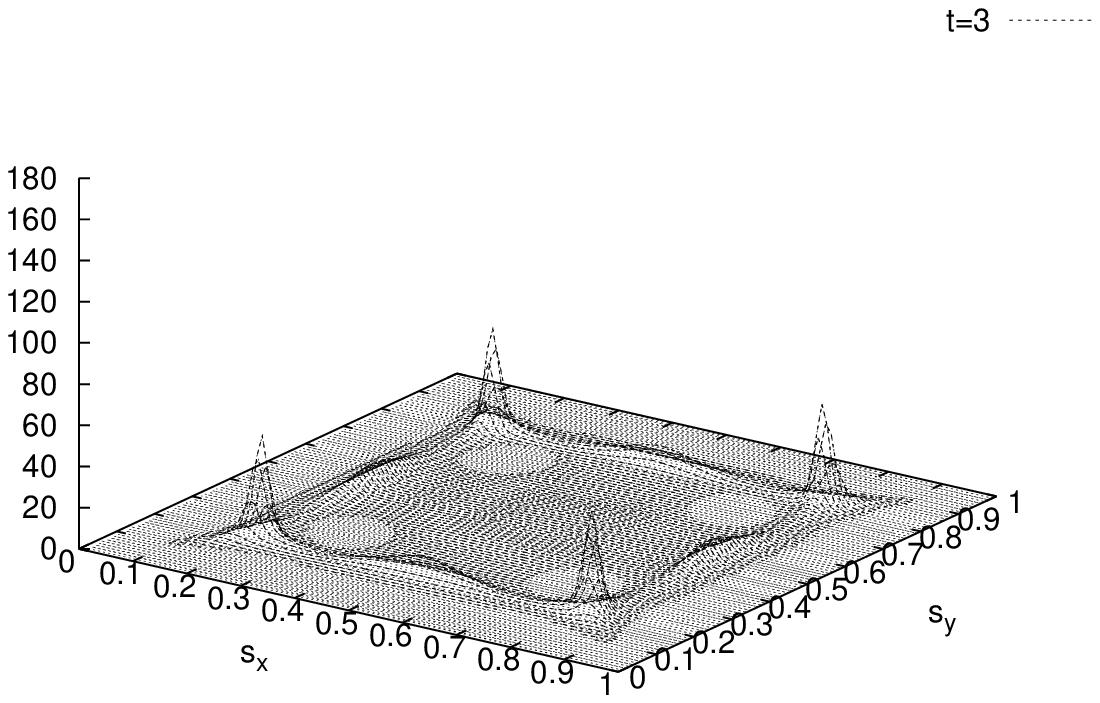, scale=0.47}
}
\centerline{
\epsfig{file=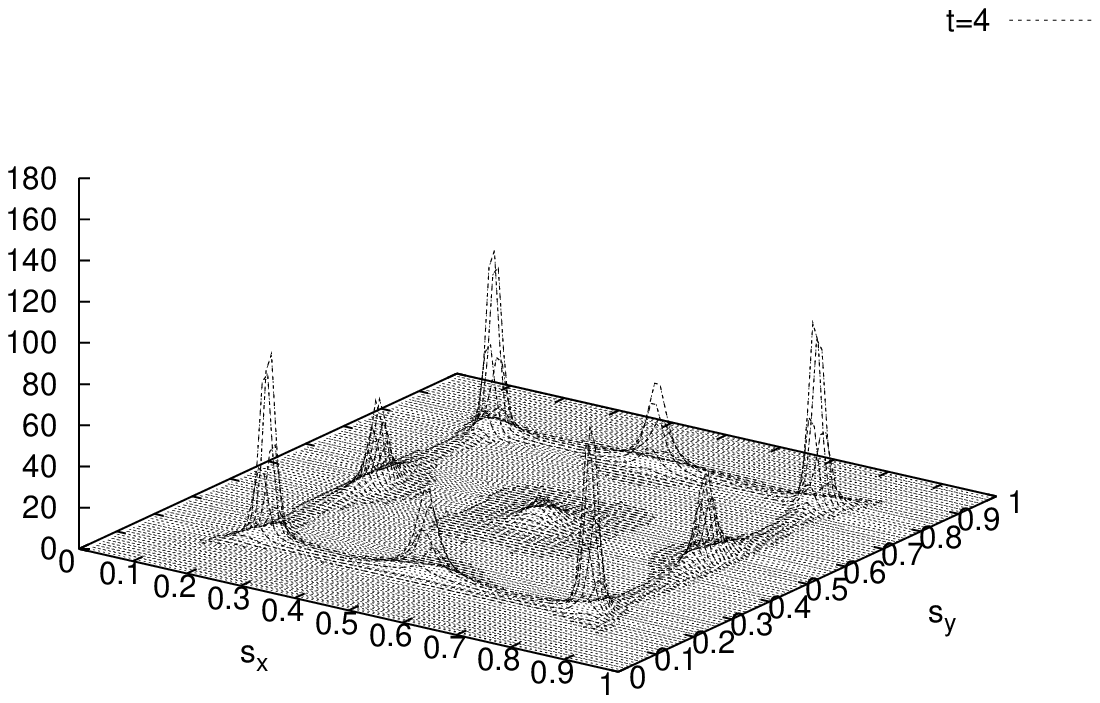, scale=0.47}
\epsfig{file=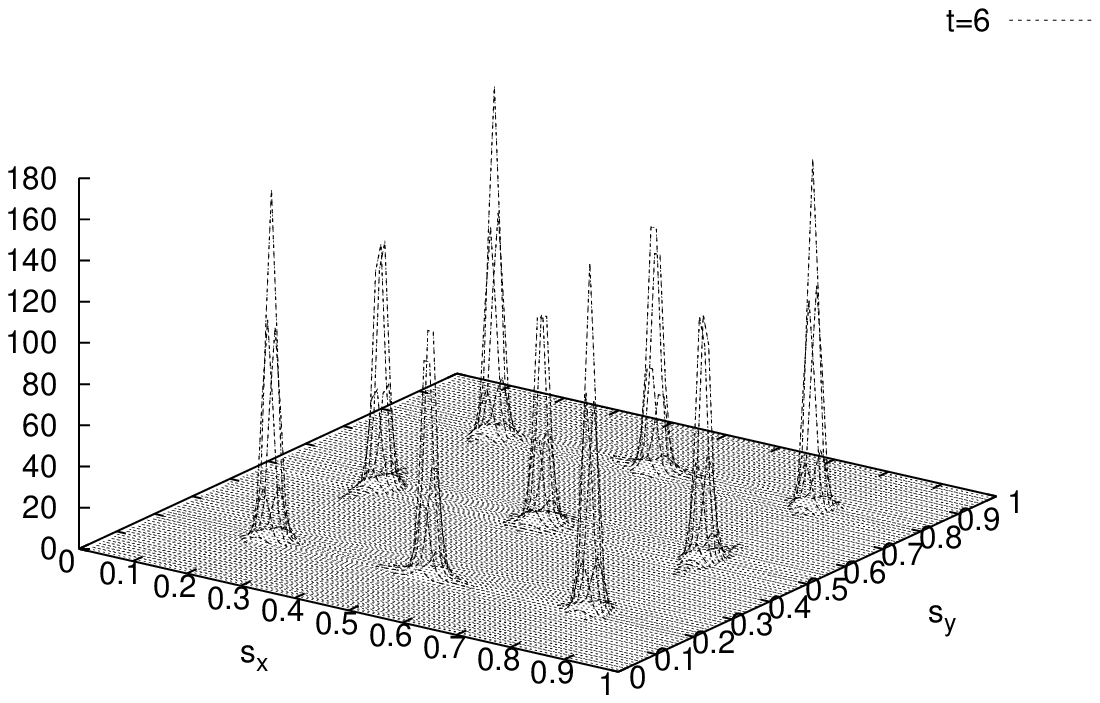, scale=0.47}
}

\fcaption{\label{fig61} As Fig. \ref{fig6} ($\epsilon$=0.15) but for circular confidence range and t=2,3,4,6.}
\end{figure}

In Fig. \ref{fig61} we take four pictures of the dynamics of the system for
$\epsilon=0.15$. The pattern looks very much the same as in the
corresponding Fig. \ref{fig6}. The number of clusters, their ordering
in opinion space and the ratio of the cluster masses are the same. 
Other trials for different values of the confidence bound
show that this is not a coincidence: the circular confidence range does not
change much the situation. In particular, the consensus threshold is only slightly higher than in the previous case, about $0.23$.
There are two reasons for that:

\begin{itemize}
\item{the surface of the circle is close to that of the square with the same linear dimension
(the ratio is $\pi/4\sim 0.8$),
so that the sets of compatible agents in the two cases considerably overlap;}
\item{the dynamics always starts from the edges of the opinion space, where the
opinion distribution is necessarily inhomogeneous, so that it is essentially the
shape of the opinion space
which rules the symmetry of the resulting opinion landscape.}
\end{itemize}

\begin{figure}[hbt]
\centerline{
\epsfig{file=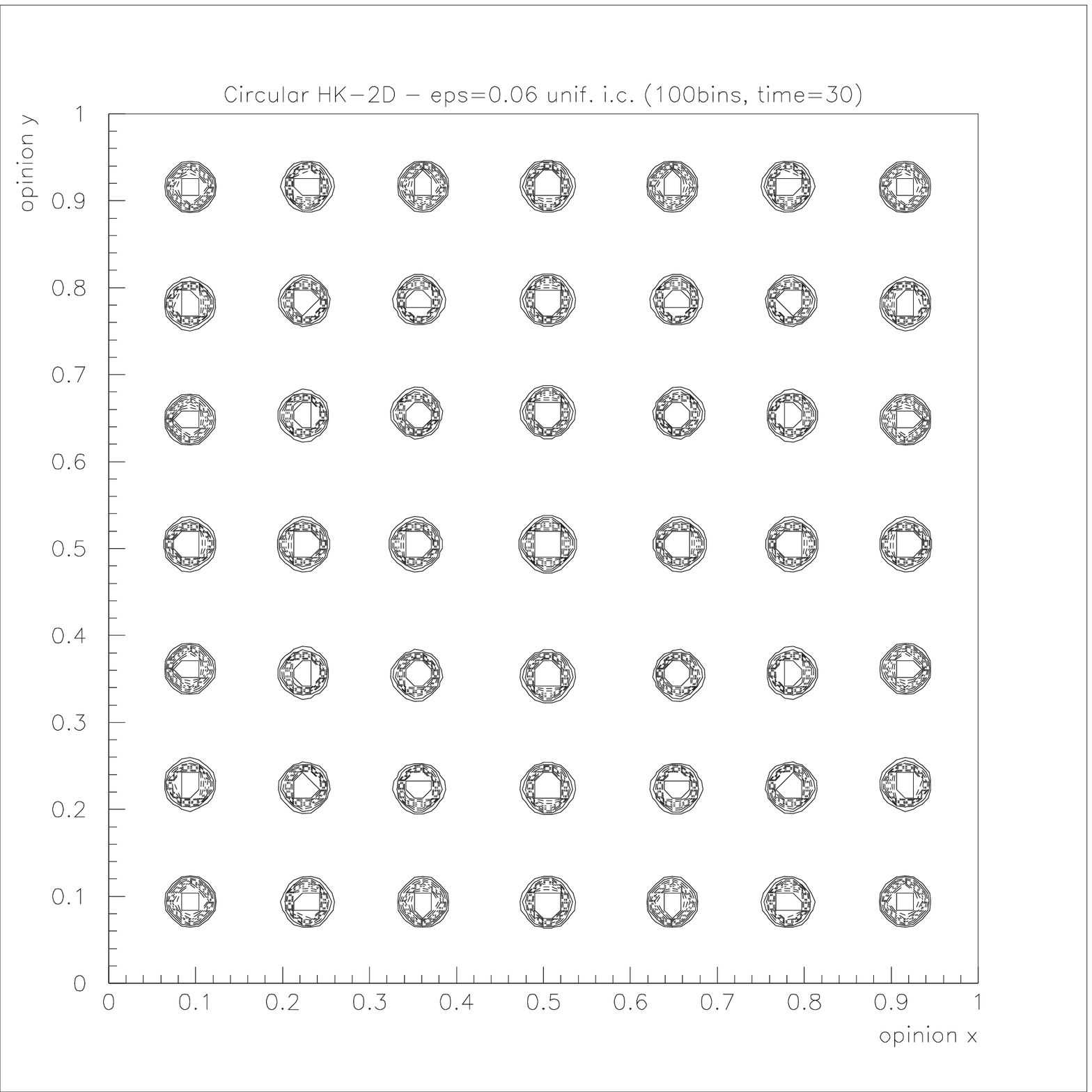, scale=0.25}
\epsfig{file=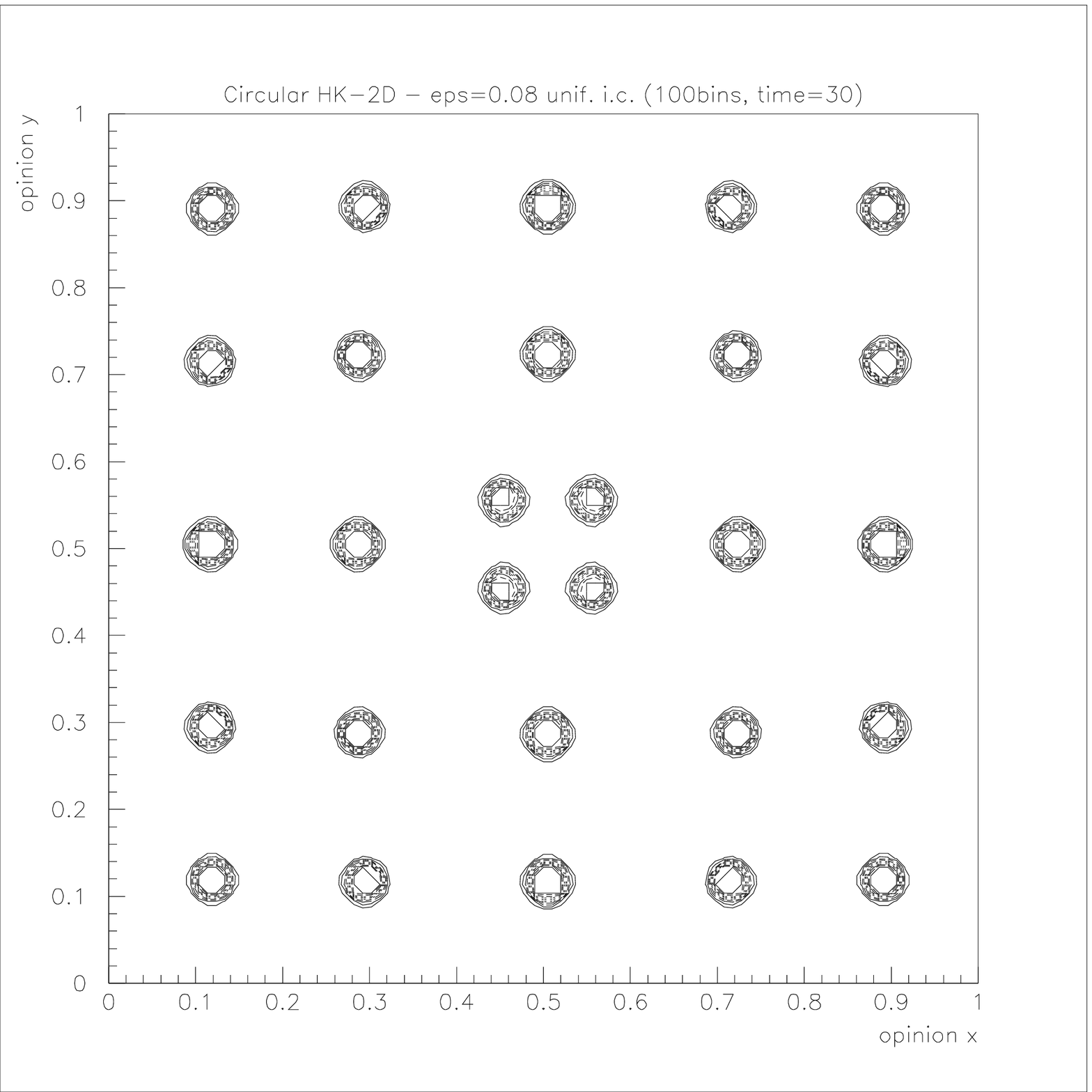, scale=0.25}
\epsfig{file=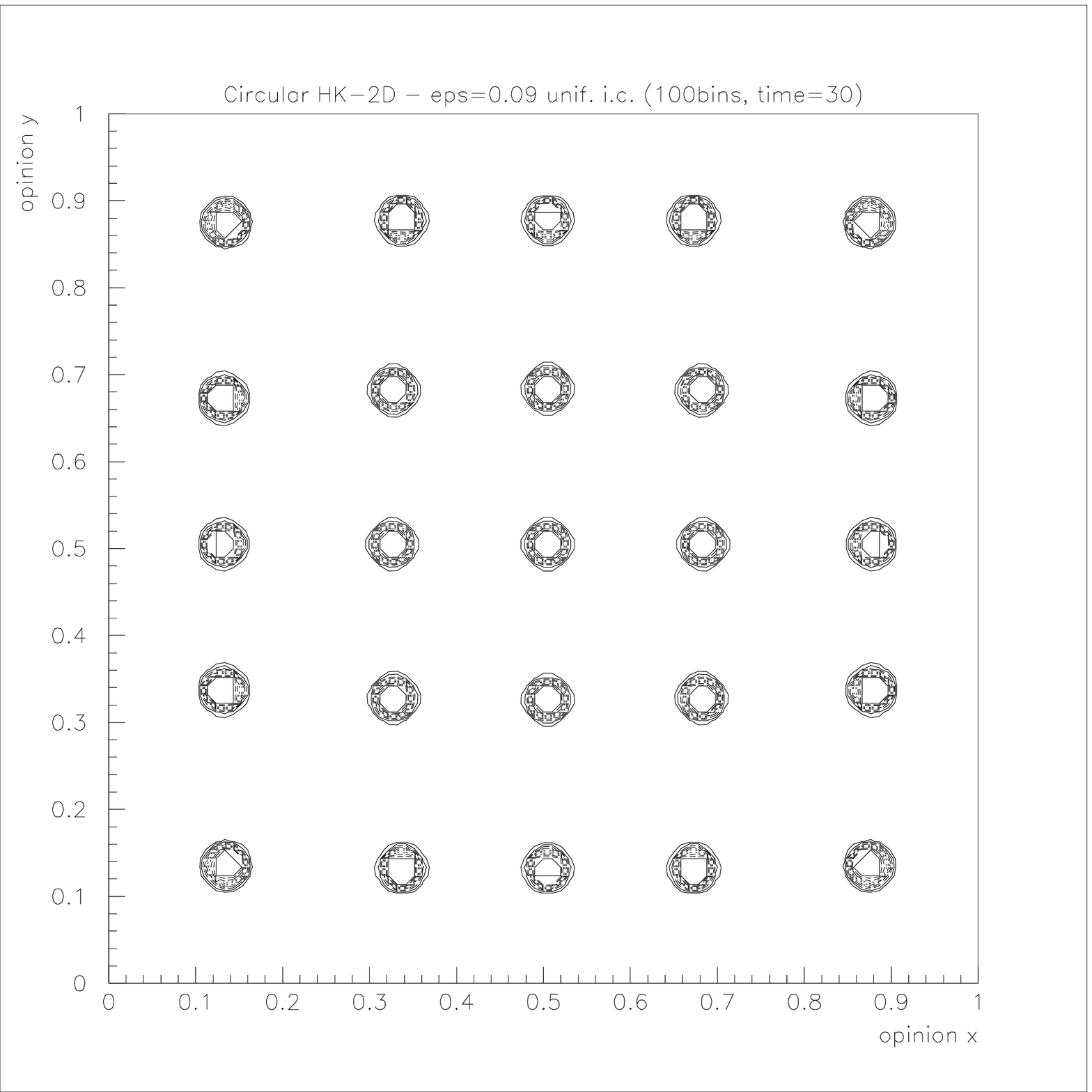, scale=0.25}
}
\centerline{
\epsfig{file=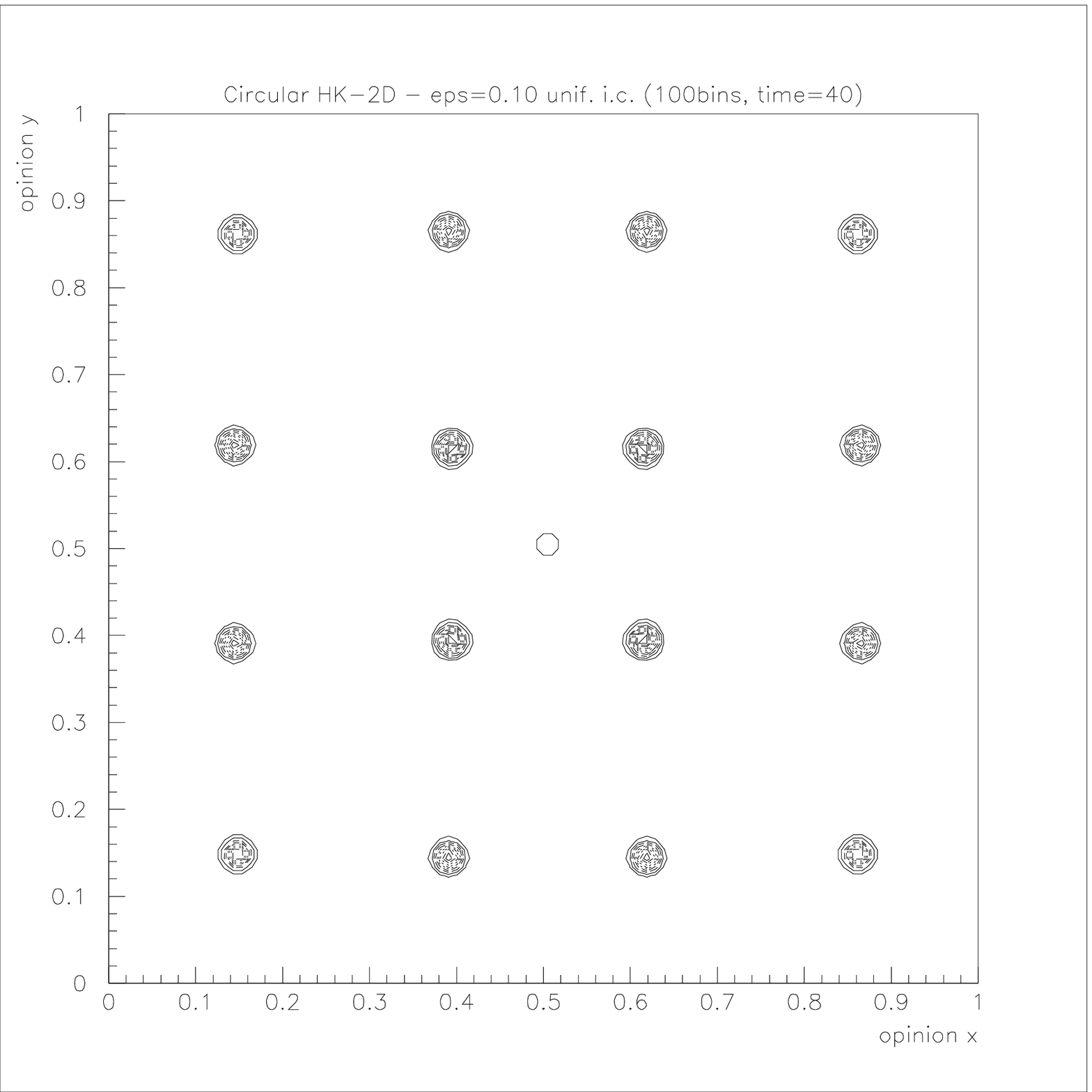, scale=0.25}
\epsfig{file=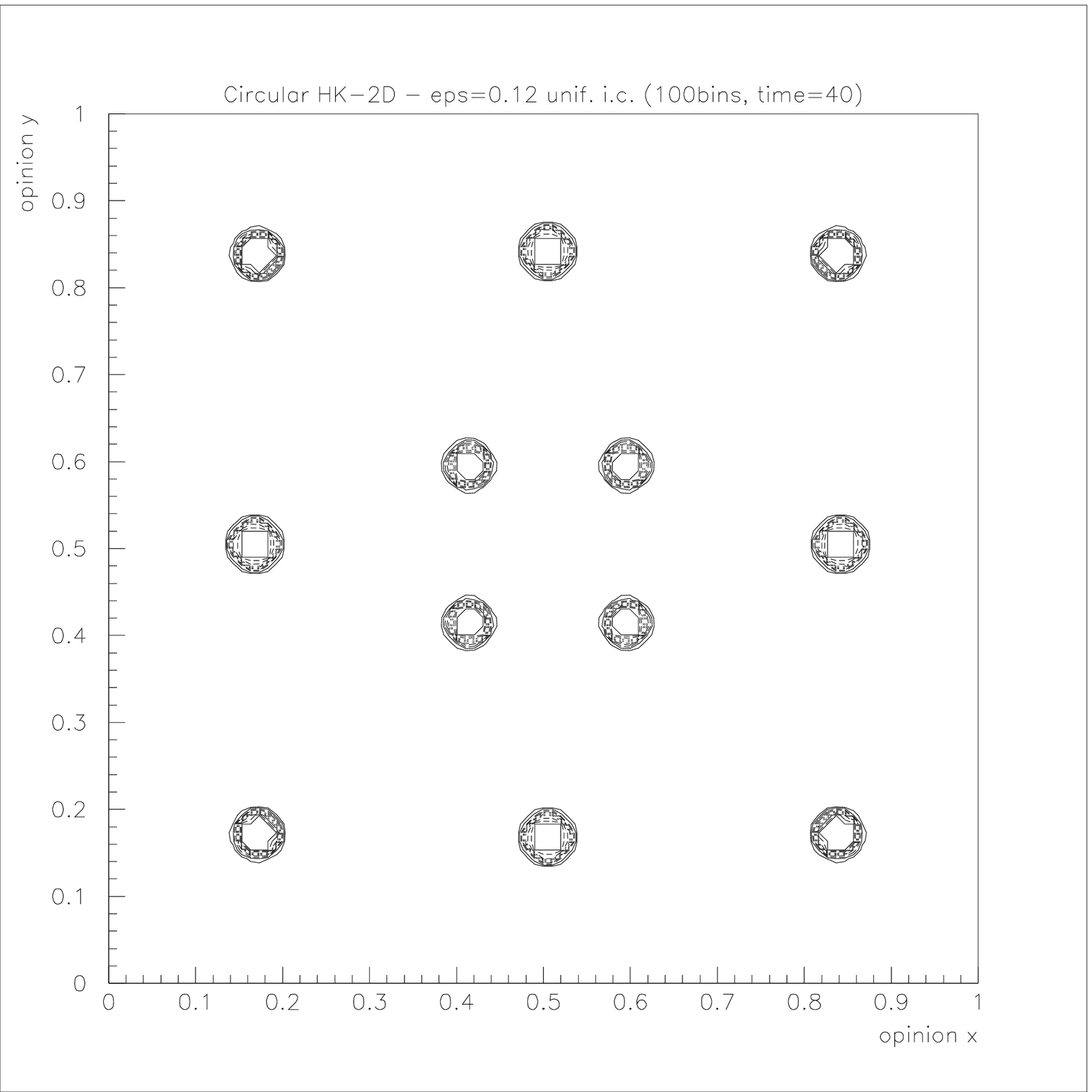, scale=0.25}
\epsfig{file=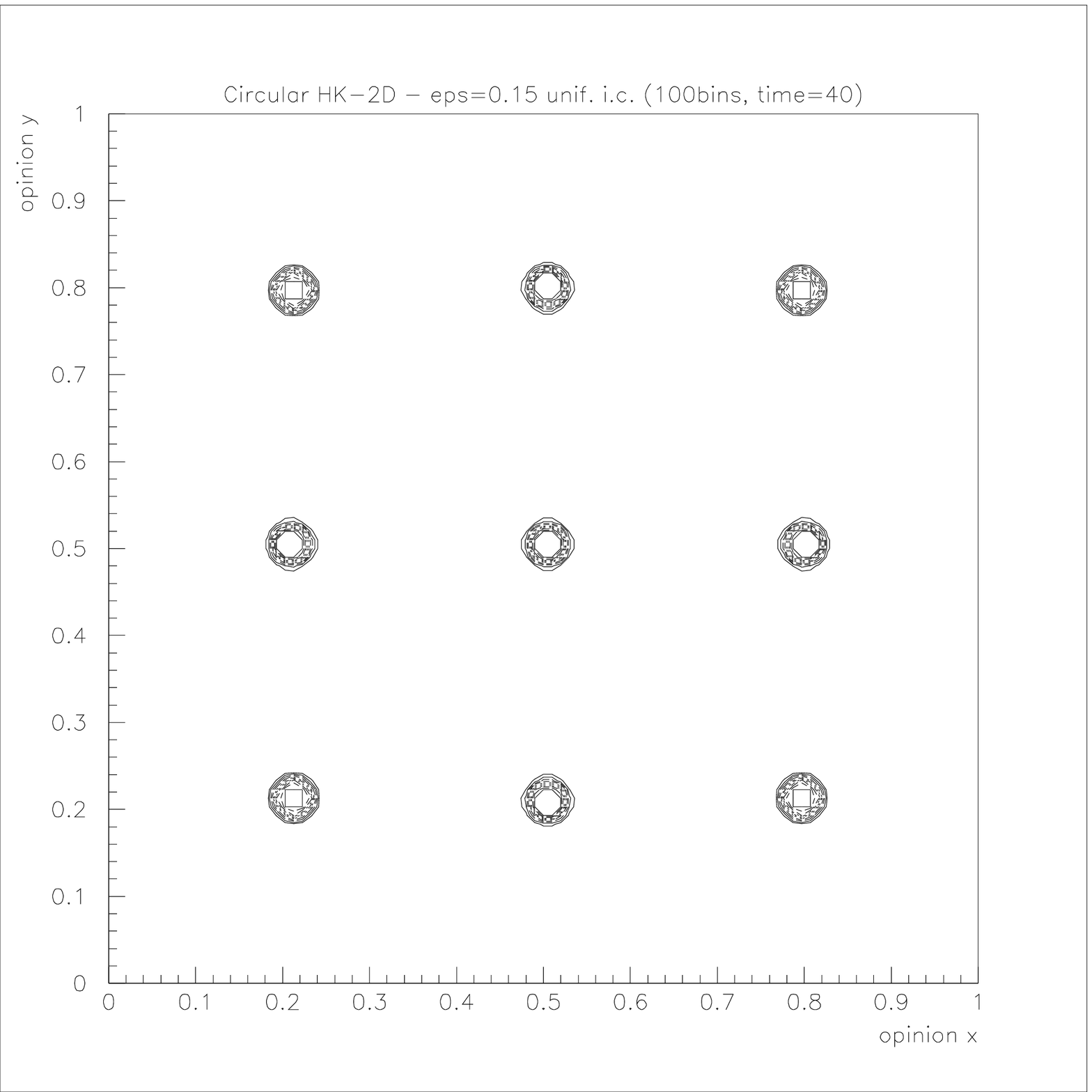, scale=0.25}
}
\centerline{
\epsfig{file=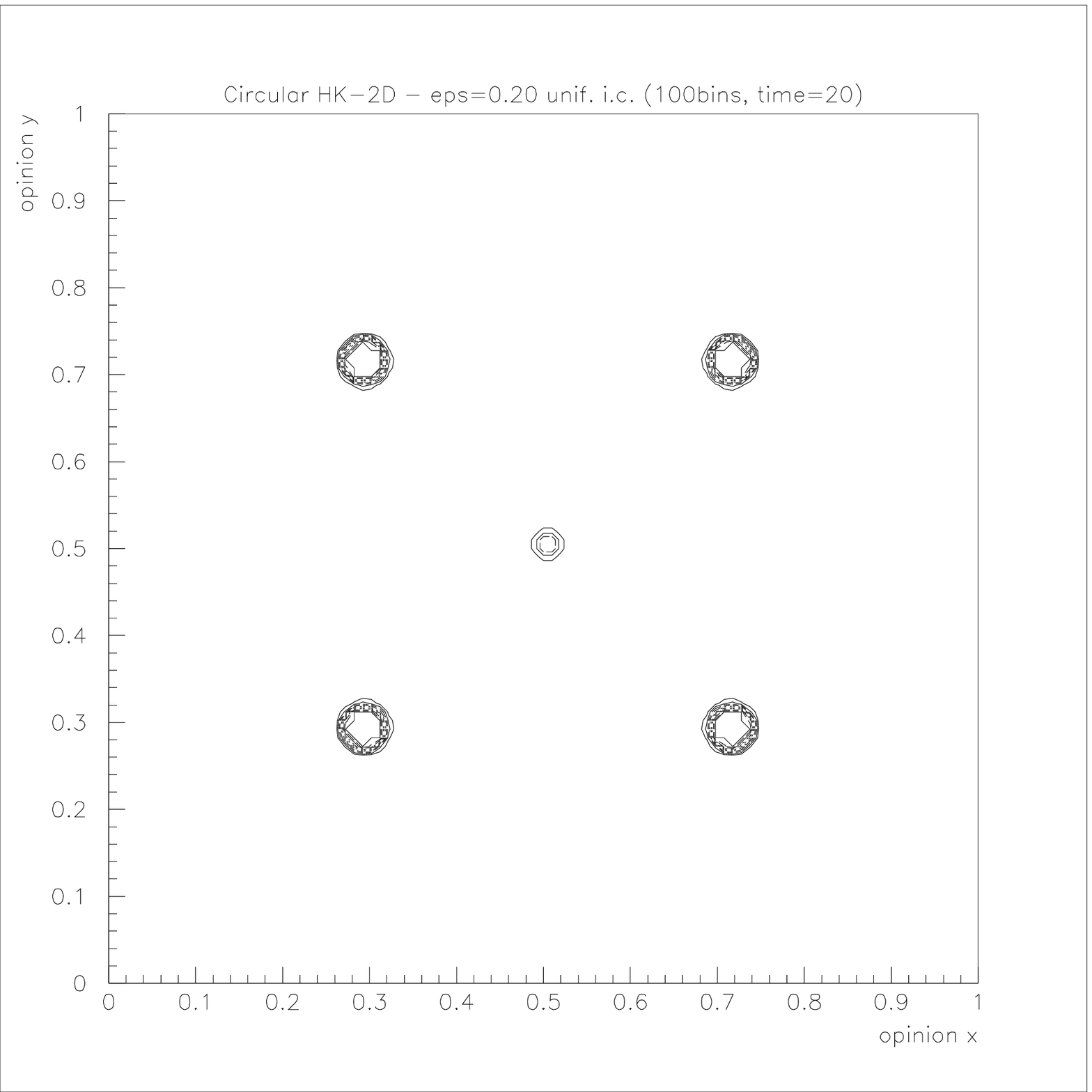, scale=0.25}
\epsfig{file=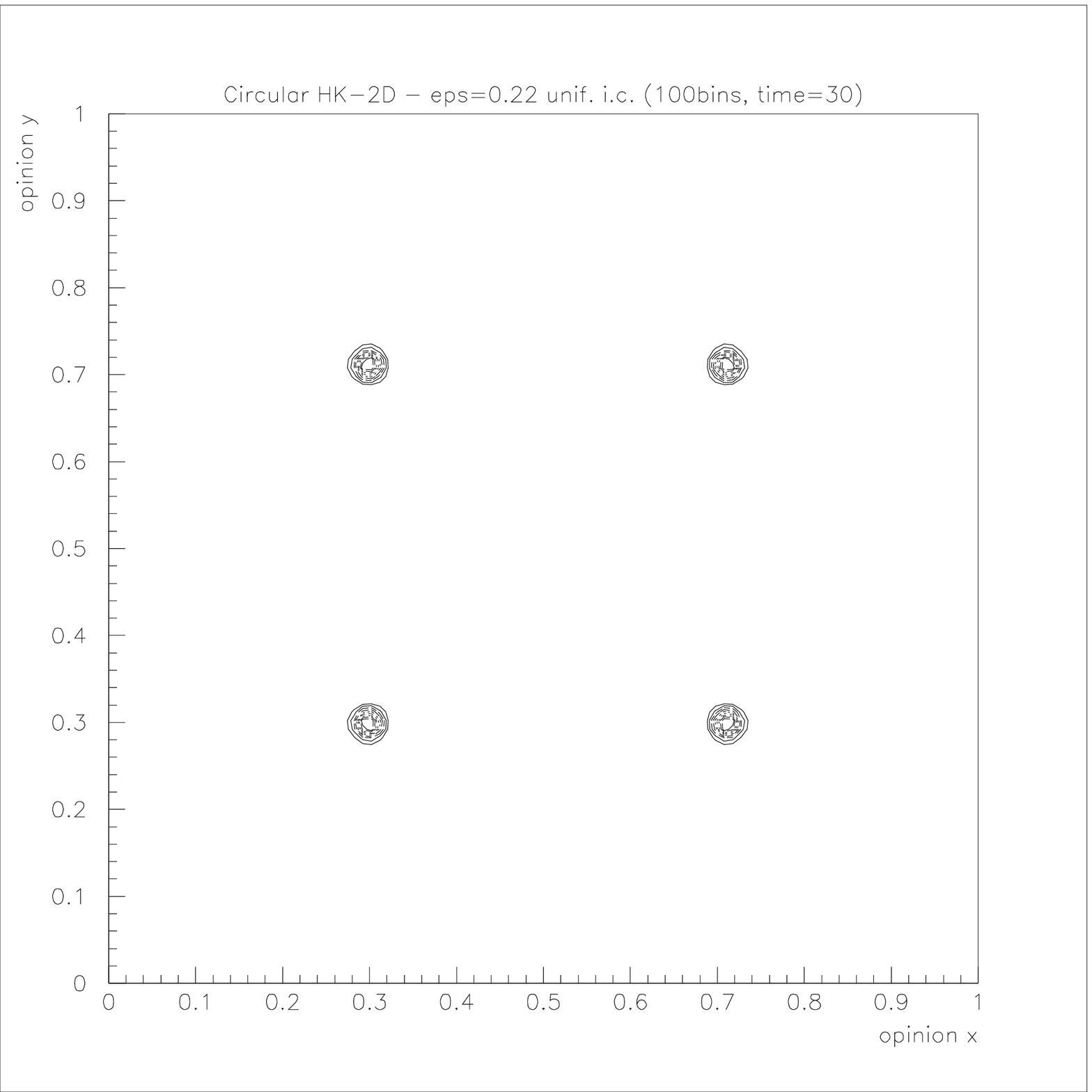, scale=0.25}
\epsfig{file=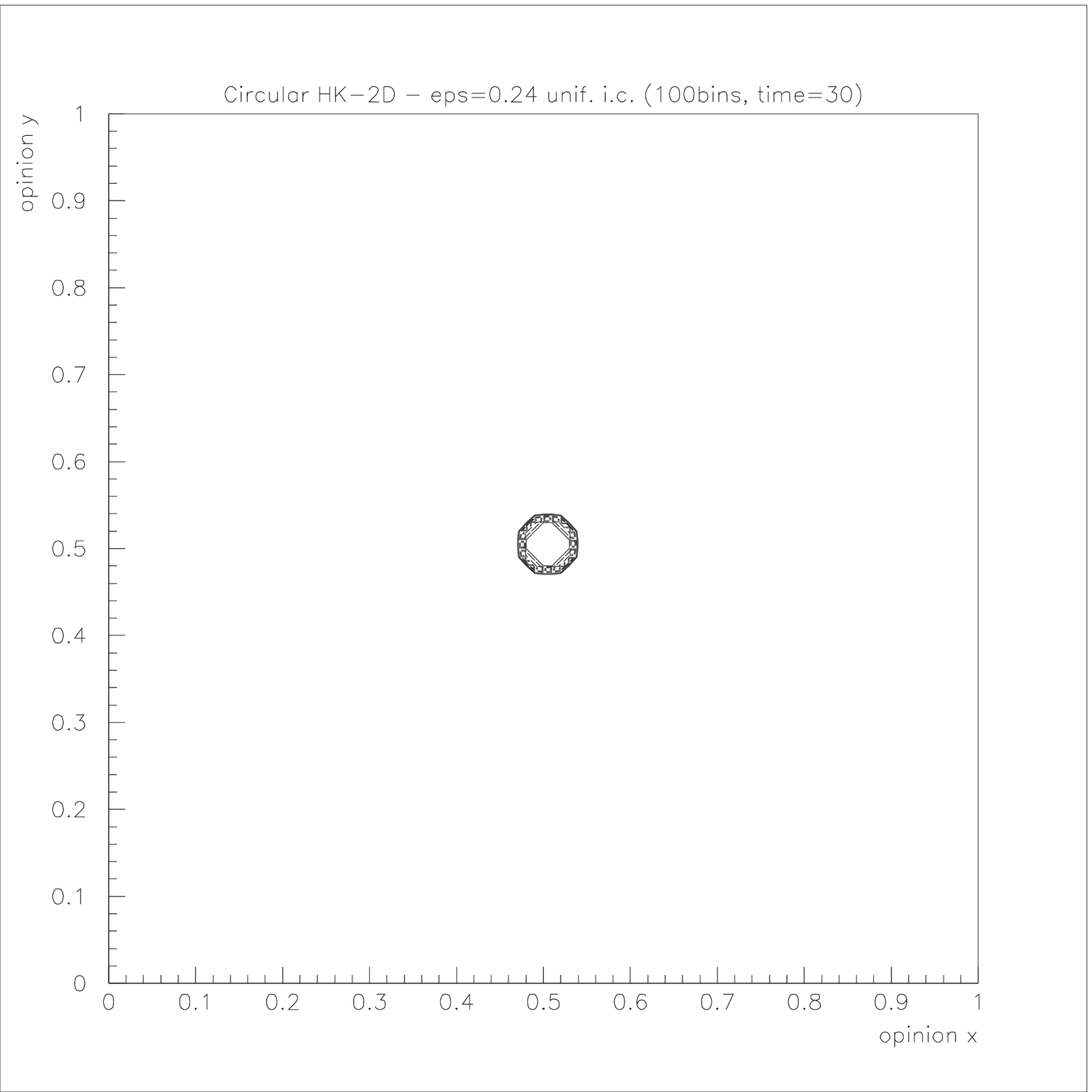, scale=0.25}
}
\fcaption{\label{figcirc}
As in Fig. \ref{figsq}, but for circular confidence range. From top left to bottom right: $\epsilon$=0.06,0.07,0.09; 
0.10, 0.12,0.15;0.20,0.22,0.24.}
\end{figure}

To have an overview of the situation, we report a series of contour plots relative to
the final stage of
the evolution at various $\epsilon$, like in Fig. \ref{figsq}. The resulting Fig. \ref{figcirc} confirms
that the clusters indeed sit on the sites of a square lattice, as we have seen above.
There are also important differences, however. Particularly striking is the occasional existence of groups of four
clusters near the center, which lie closer to each other as compared to the other clusters (see, for instance,
the patterns corresponding to
$\epsilon=0.08$ and $\epsilon=0.12$). Moreover, as we have seen for the
case of the squared confidence range, sometimes smaller clusters survive
on (some) sites of the dual lattice, especially at the center of the opinion space
(like for $\epsilon=0.10$ and $\epsilon=0.20$).

\begin{figure}[hbt]
\centerline{
\epsfig{file=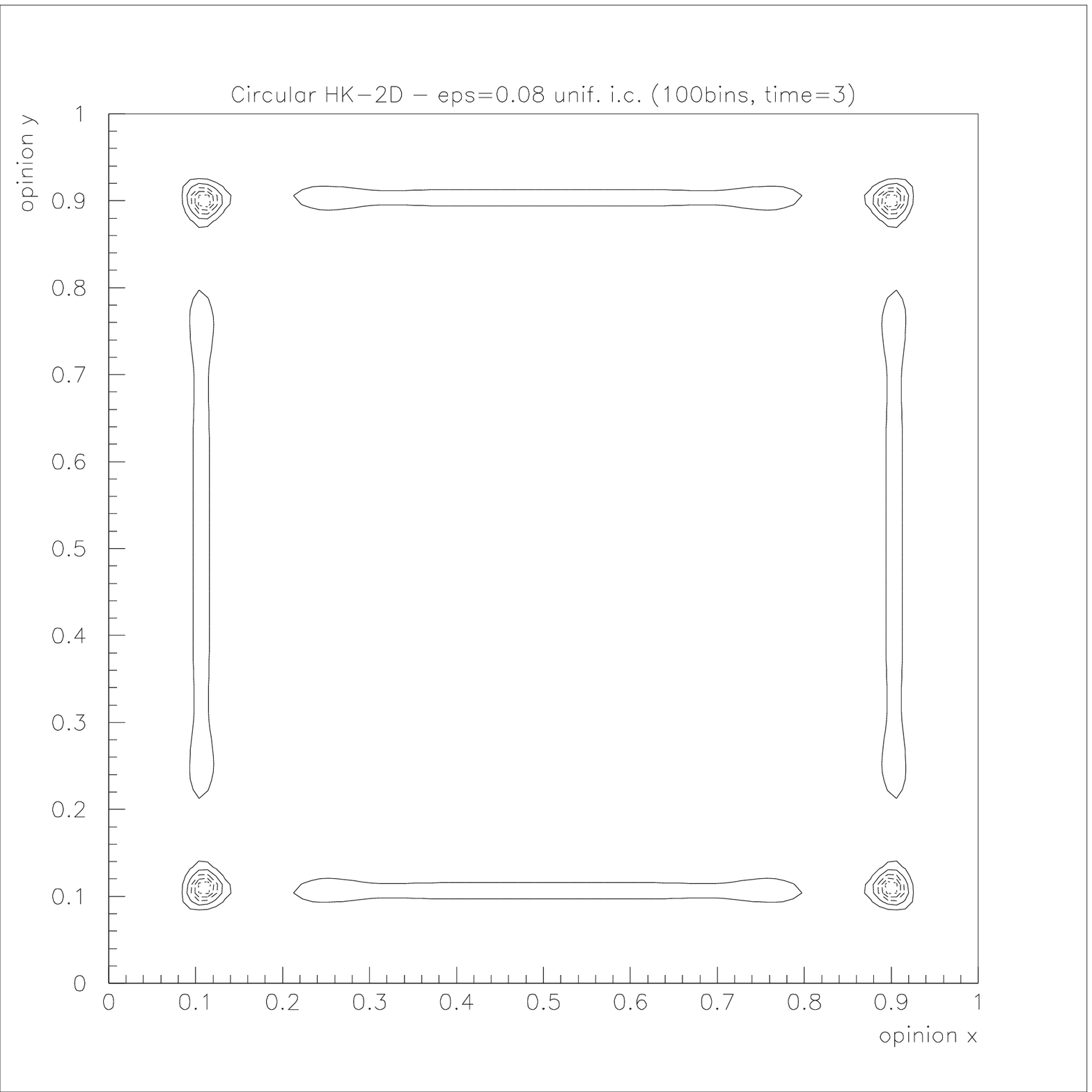, scale=0.25}
\epsfig{file=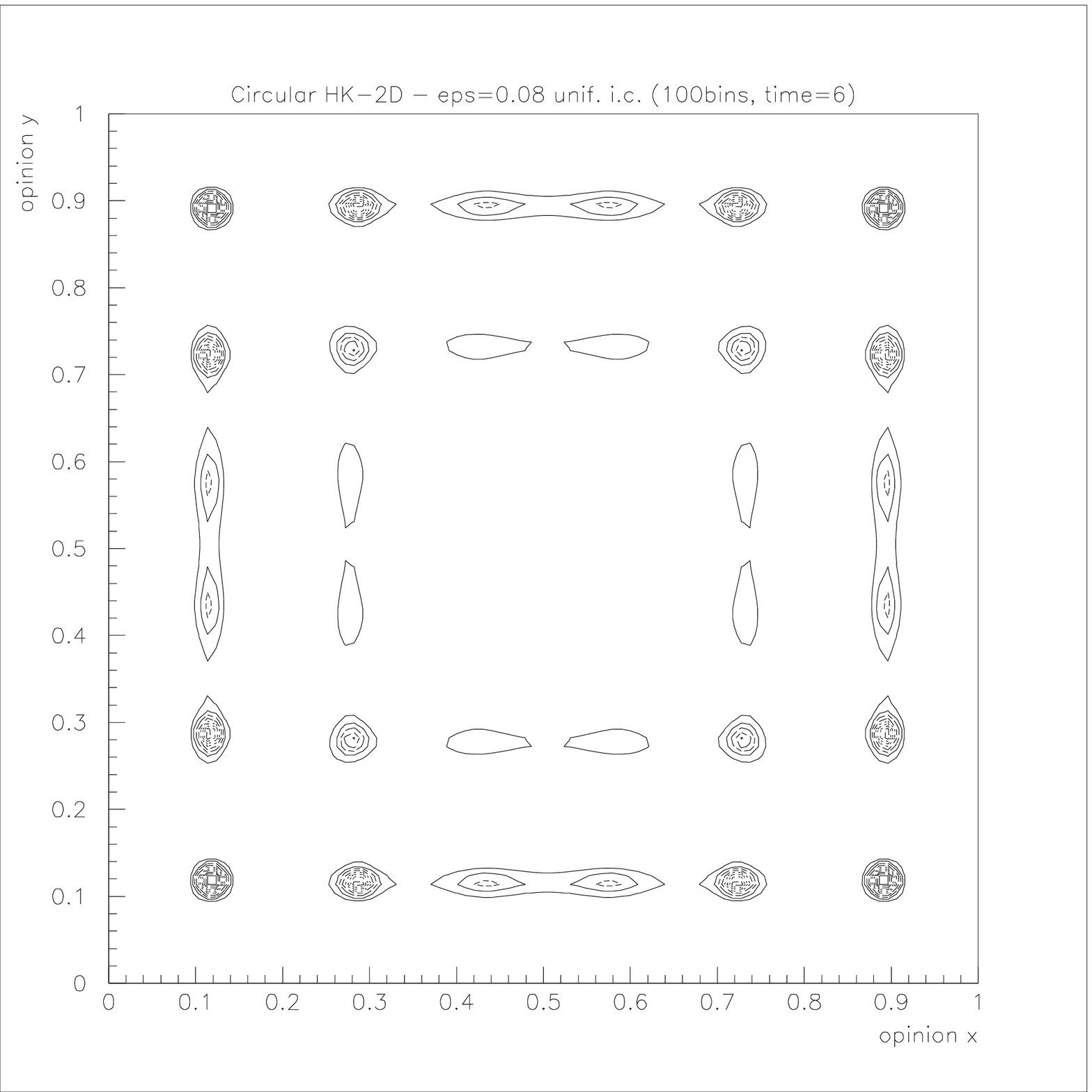, scale=0.25}
\epsfig{file=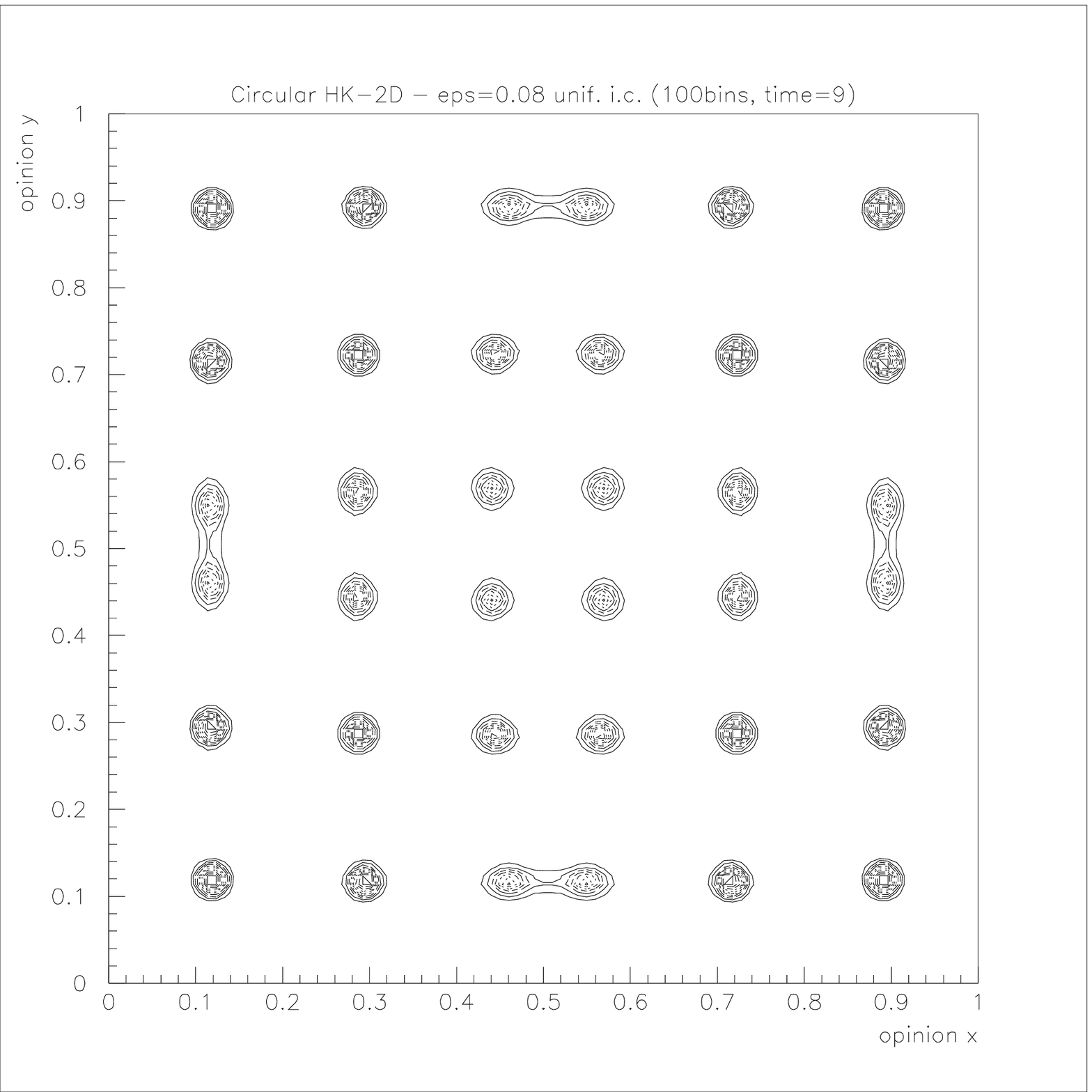, scale=0.25}
}
\centerline{
\epsfig{file=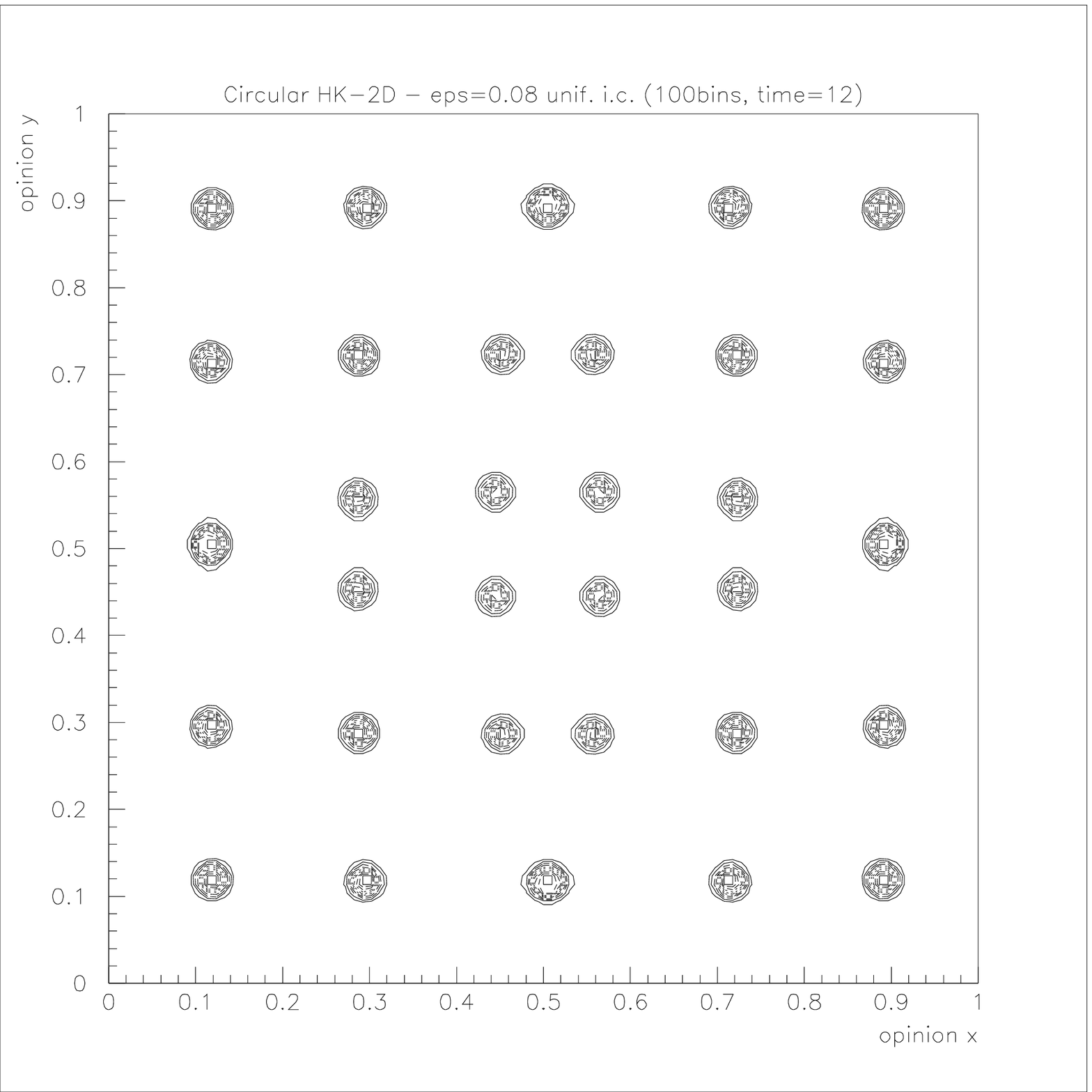, scale=0.25}
\epsfig{file=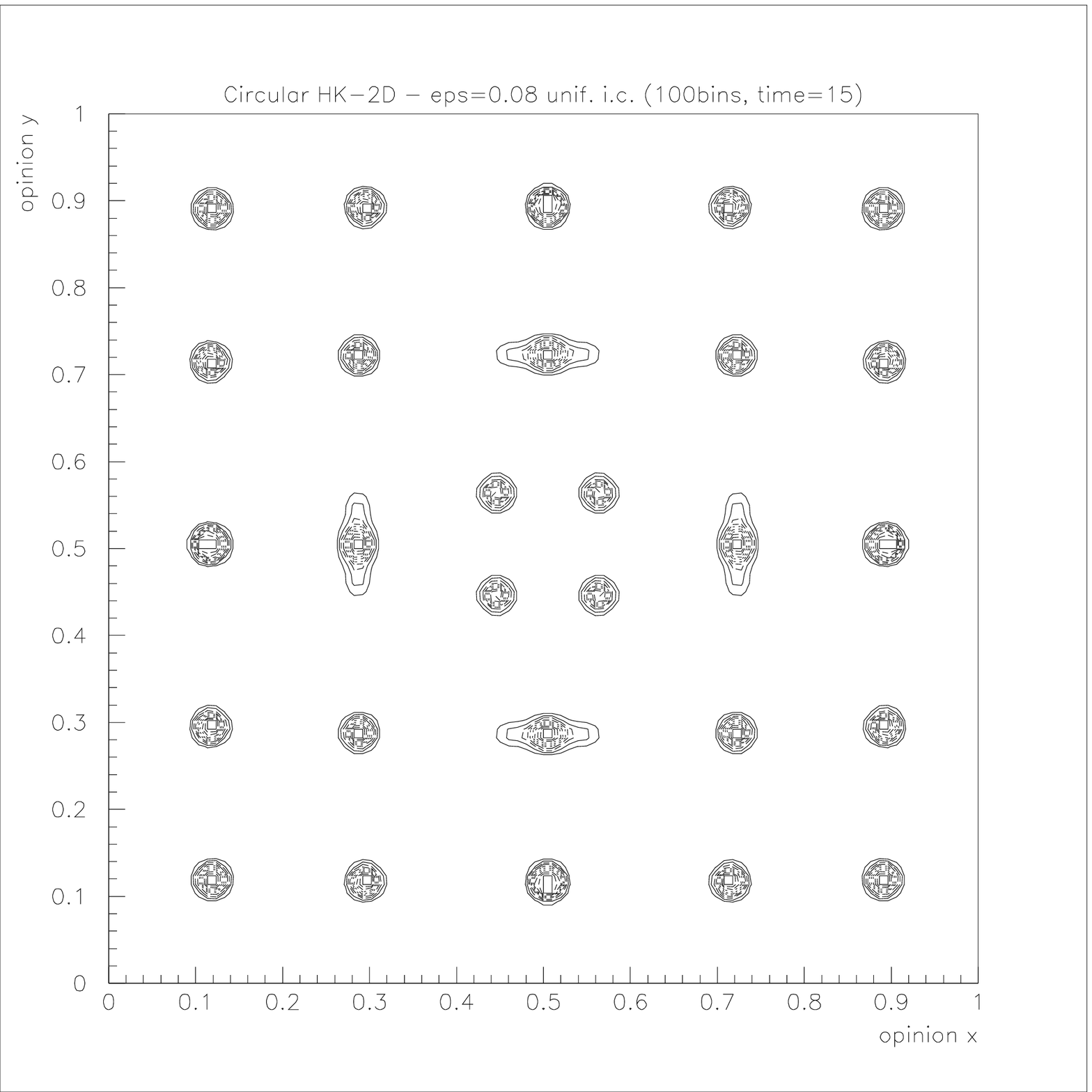, scale=0.25}
\epsfig{file=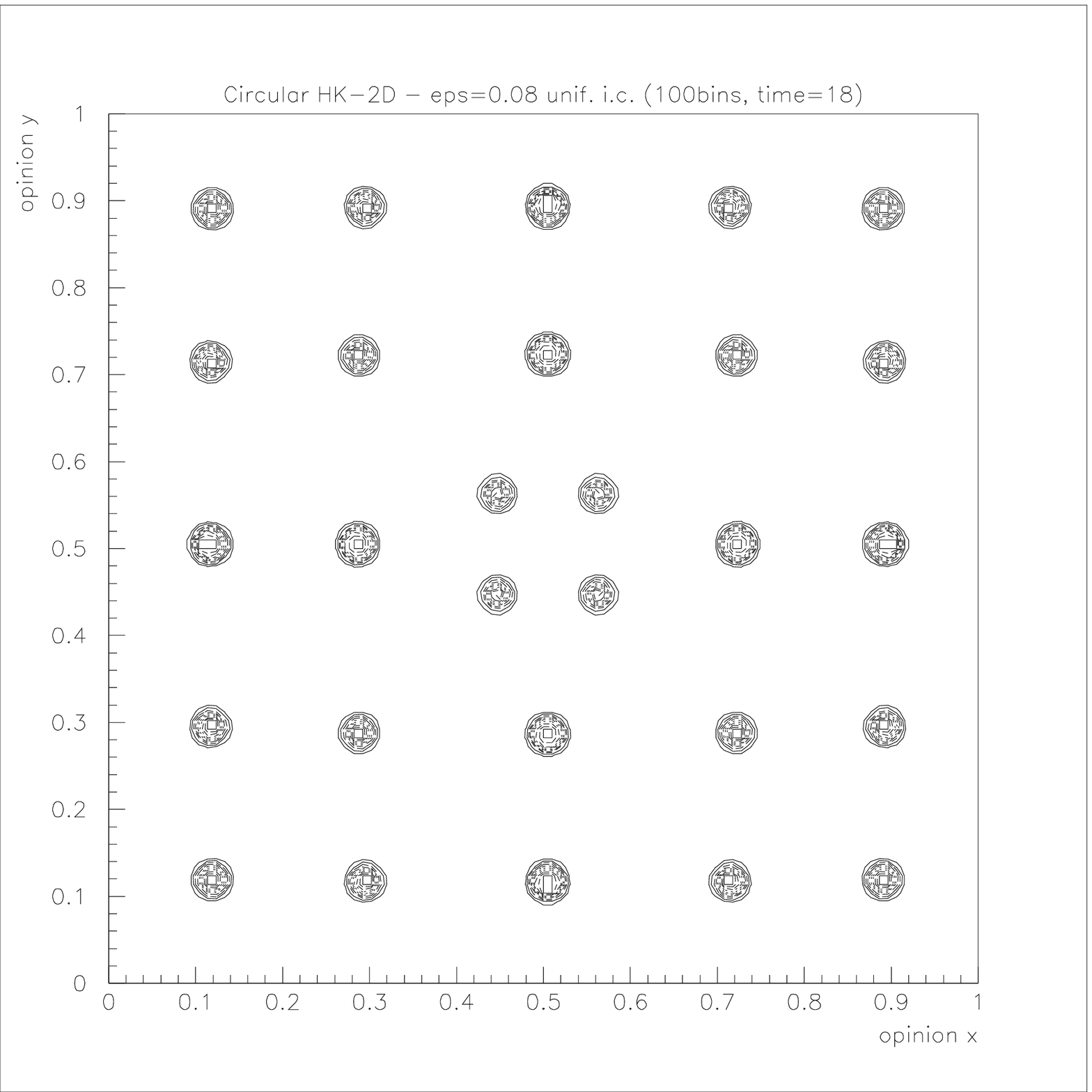, scale=0.25}
}
\vspace*{0.5cm}
\fcaption{\label{figlast}
Six snapshots show the temporal evolution of cluster formation
and merging for circular confidence range and $\epsilon=0.08$.
From top left to bottom right: t=3, 6, 9; 12, 15, 18.}
\end{figure}

It is also interesting to show the cluster formation in the case in which a group of four clusters
near the center appears and remains stable in time.
In Fig.\ref{figlast} a sequence of six contour plots calculated at different times is
shown for $\epsilon=0.08$. As one can see, the first snapshot confirms the previous
statement that the symmetry of the opinion configurations is fixed by the shape of
the opinion space. Going on, one can observe the progressive merging of the pairs of
clusters with reciprocal distance less than the confidence bound radius. Finally,
in the last picture, the survival of the four "anomalous" central clusters indicates
that these clusters lie exactly at the border of the confidence range and this is clearly
an effect of the circular shape of the confidence range (with a squared range, these four
clusters would be attracted towards the center). We could also interprete it as a typical
effect of the interdependence between the two components of the opinion vector.

\section{Conclusions}

We have extended the continuum opinion dynamics of Krause-Hegselmann
to the case in which the opinion is not just a scalar but a vector with real-valued components.
The extension is straightforward, with some freedom in the choice of the shape
of the confidence range and the opinion space.
Here we took a square for the opinion space and
a square and a circle for the confidence range.
We investigated a community where everybody talks to everybody else,
and analyzed the dynamics and the final opinion configurations by solving numerically
a rate equation. We found that if we project the final opinion configurations on any component,
the number of clusters is essentially
the same as in the one-dimensional opinion dynamics. The consensus threshold is
slightly larger for the circular confidence range because the area spanned by the circle is smaller than
that spanned by the square with the same linear dimension, but the two values are close to each other
and to the one-dimensional consensus threshold. The structure formed by the centers of the final
opinion clusters is a regular square lattice, but for special values of the confidence bound peculiar patterns
also occur: survival of small clusters on the sites of the dual lattice, merging of the innermost four clusters
into a large central one, existence of a compact group of four clusters near the center.

\clearpage

Summing up, we have found that vector opinion dynamics induces no significative variation in the
evolution of the system that cannot be deduced by combining the results of the simple one-dimensional
dynamics of the single opinion components.
We studied here just the case of bidimensional opinions, but we do not expect
big changes for a higher number of opinion components. However, we should not forget that we
investigated a particularly simple model on a complete graph starting from a uniform opinion distribution,
and this is at best only a zero$th$-order approximation of what happens for real systems.

S. F. acknowledges the financial support of the Volkswagen Foundation.

\nonumsection{References}

\end{document}